\documentclass[pra,twocolumn,nofootinbib,longbibliography,superscriptaddress,9pt]{revtex4-1}

\usepackage[utf8]{inputenc}
\usepackage[T1]{fontenc} \usepackage{lineno}
\usepackage{graphicx}\usepackage{dcolumn}
\usepackage{bm}\usepackage{commath}
\usepackage{latexsym}\usepackage{braket}
\usepackage{amssymb}\usepackage{amsmath}
\usepackage{mathtools}\usepackage{multirow}
\usepackage{capt-of}
\usepackage{color}
\usepackage{ragged2e}

\makeatletter
\newcommand*{\rom}[1]{\expandafter\@slowromancap\romannumeral #1@}
\makeatother
\newcommand{\D}{\mathrm{d}}
\usepackage[colorlinks,breaklinks,linkcolor={blue},citecolor={magenta},urlcolor={blue}]{hyperref}

\begin{document}

\title{Modeling beam propagation in a moving nonlinear medium}

\author{Ryan Hogan$^\dagger$}
\email[Corresponding author: ]{ryan.hogan.j@gmail.com}
\affiliation{Department of Physics, University of Ottawa, Ottawa ON K1N 6N5, Canada}

\author{Giulia Marcucci$^\dagger$}
\affiliation{Department of Physics, University of Ottawa, Ottawa ON K1N 6N5, Canada}
\affiliation{School of Physics \& Astronomy, University of Glasgow, Glasgow G128QQ, United Kingdom}

\author{Akbar Safari}
\affiliation{Department of Physics, University of Ottawa, Ottawa ON K1N 6N5, Canada}
\affiliation{Department of Physics, University of Wisconsin-Madison, 1150 University Avenue, Madison, WI, 53706, USA}

\author{A. Nicholas Black}
\affiliation{Department of Physics and Astronomy, University of Rochester, Rochester, NY 14627, USA}

\author{Boris Braverman}
\affiliation{Department of Physics, University of Ottawa, Ottawa ON K1N 6N5, Canada}

\author{Jeremy Upham}
\affiliation{Department of Physics, University of Ottawa, Ottawa ON K1N 6N5, Canada}

\author{Robert W. Boyd}
\affiliation{Department of Physics, University of Ottawa, Ottawa ON K1N 6N5, Canada}
\affiliation{Institute of Optics, University of Rochester, Rochester, NY 14627, USA}

\date{\today}
\begin{abstract}
Fully describing light propagation in a rotating, anisotropic medium with thermal nonlinearity requires modeling the interplay between nonlinear refraction, birefringence, and the nonlinear group index. Incorporating these factors into a generalized nonlinear Schrödinger equation and fitting them to recent experimental results reveals two key relationships: the photon drag effect can have a nonlinear component that is dependent on the motion of the medium, and the temporal dynamics of the moving birefringent nonlinear medium create distorted figure-eight-like transverse trajectories at the output. The beam trajectory can be accurately modelled with a full understanding of the propagation effects. Efficiently modeling these effects and accurately predicting the beam’s output position has implications for optimizing applications in velocimetry and beam-steering. Understanding the roles of competitive nonlinearities gives insight into the creation or suppression of nonlinear phenomena like self-action effects. 
\end{abstract}
\maketitle
\def\thefootnote{$\dagger$}\footnotetext{These authors contributed equally to this work}\def\thefootnote{\arabic{footnote}}
\section{Introduction\label{sec:intro}}
Light propagation in a moving medium is subject to photon drag. Drag was first predicted by Fresnel~\cite{fresnel1818} and later experimentally proved by Fizeau~\cite{fizeau1860xxxii}. Depending on the direction of medium motion relative to the optical path, light drag can change the speed of light in the longitudinal direction or shift the beam in the transverse direction. These changes are typically minute and require sensitive measurements to be observed~\cite{sanders1988measurement}. However, it has been shown that a large group index can enhance both longitudinal~\cite{2016Safari} and transverse~\cite{franke2011rotary} light drag. Moreover, the light drag effect is linearly proportional to the speed of the moving medium. In the case of the transverse drag, fast transverse motion can be achieved using the tangential component of rotational motion far from the axis of rotation. In some cases, rotation is also helpful in producing slow light effects~\cite{franke2011rotary}. However, rotation can add complexity, particularly when considering birefringent media, which requires additional considerations. Therefore, modeling light propagation subject to large transverse shifts must account for the rotation rate, birefringence, and large group indices. Moreover, if the light is intense, the impact of any optical or thermal nonlinear response, acting both locally and nonlocally, must also be considered. Transverse shifts from photon drag have been modelled as a linear effect ~\cite{fizeau1860xxxii,fresnel1818,franke2011rotary,piredda2007slow,2003Carusotto}. However, when thermal and optical nonlinearities become significant, we must incorporate the nonlinear response effect on the group index, including the different time scales over which they will impact the direction and the magnitude. 

In this work, we introduce a general theory to describe the interaction of linearly polarized light with a rotating birefringent nonlocal nonlinear medium. Our model considers rotation, birefringence, and nonlinear refraction, and by incorporating the nonlinear contributions to the material’s group index, we extend the linear photon drag effect to the nonlinear regime. All effects can be incorporated using an intensity- and rotation speed-dependent dielectric tensor, then we develop coupled generalized nonlinear Schrödinger equations for the ordinary (o-) and extraordinary (e-) beams to fully describe the linear and nonlinear dynamics. As a result, we can plot the distorted beam trajectories through the medium and their transverse shifts. 

Our theoretical work applies to any light propagation in a rotating, linear, or nonlinear medium~\cite{boccia2009tunable} and supports our previous experimental work~\cite{hogan2023beam}. Since the trajectories are tracked and controllable, our work has implications for applications in beam-steering~\cite{zhou2021optical}. Moreover, modeling the polarization response due to dielectric tensor could lead to manipulated propagation of vector beams~\cite{allen1963new}.



\section{The photon drag effect \label{sec:0}}

As light travels through a moving medium, the speed of light with respect to the laboratory frame changes, producing light drag. However, analogously to how the different phase velocities of the constituent frequencies of an optical pulse determine its group velocity, the different phase shifts of the constituent momentum components of an optical beam determine its path. Medium movement can be either along the propagation direction, producing an optical phase shift and longitudinal drag~\cite{2016Safari}, or perpendicular to propagation, inducing transverse drag~\cite{2003Carusotto}.

Focusing on the transverse drag case (Fig. 1), Carusotto et. al.~\cite{2003Carusotto} derives the transverse beam deflection ($\Delta y$) for monochromatic, collimated light interacting with an isotropic, lossless, dispersive, linear medium of length L and in motion with constant speed, $v$.



Upon entering the moving medium at normal incidence, the beam deflects from its direction of propagation by some angle $\theta$, as determined by its phase index $n_0$ and group index, $n_g$, 

\begin{figure}[t!]
\begin{center}
\includegraphics[width=.99\columnwidth]{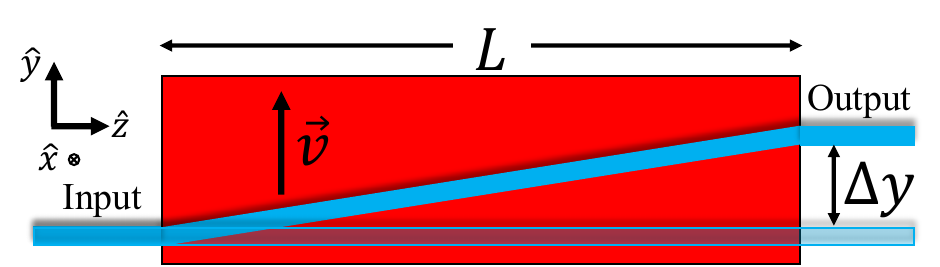}
\end{center}
\caption{
{\bf Schematic of the linear, transverse photon drag effect.} A beam of light passing through an isotropic, lossless medium of length $L$, moving transversely with speed $v$, is laterally shifted by $\Delta y$, given in Eq.~\eqref{eq:lindrag}.
\label{fig:1}}
\end{figure}

\begin{equation}
\tan\theta=\frac{v}{c}\left(n_g-\frac{1}{n_0}\right),
\label{eq:dragangle}
\end{equation}
where $c$ is the speed of light in a vacuum. 
Thus, the transverse shift describing the magnitude of the photon drag effect, $\Delta y$, is
\begin{equation}
\Delta y=\frac{vL}{c}\left(n_g-\frac{1}{n_0}\right),
\label{eq:lindrag}
\end{equation}
where $L$ is the medium length. The group index can become large in certain media in the presence of intense laser beams ($|n_g|\approx10^6$~\cite{piredda2007slow,https://doi.org/10.48550/arxiv.2301.13300}), and thus the linear photon drag effect can extend to a nonlinear regime. Therefore, we must consider the lowest-order nonlinear corrections to the indices $n_0$ and $n_g$. The corrections take the form $\Delta n = n_2 I$ arising from an instantaneous Kerr-type nonlinearity where $n_2$ is the nonlinear refractive index, and $I$ is the input beam intensity. The phase and group indices then become 
\begin{equation}
\begin{aligned}
&n_{0}^{NL}= n_0+n_2 I,\\
&n_{g}^{NL}=n_g^0+n_2^g I,  
\end{aligned}
\label{eq:groupindexfinal}
\end{equation}
where $n_g^0$ is the linear group index, $n_2^g$ is the nonlinear group index
\begin{equation}
n_2^g=\left(n_2 +\omega_0 \left(\frac{\D n_2}{\D\omega}\right)_{\omega_0}\right).
\end{equation}
Here, $n_g^{NL}$ represents the overall change of the group index, including the Kerr-like nonlinear response. Substituting $n_{g}^{NL}$ in Eq.~\eqref{eq:lindrag}, we find the transverse shift including nonlinear photon drag as
\begin{equation}
\Delta{y_{NL}} = L \tan(\theta_{NL}) = \frac{L v}{c}\left(n_{g}^{NL}-\frac{1}{n_0}\right),
\label{eq:dragnonlin}
\end{equation}
where $n_2 I \ll n_0$. When the magnitude of the group index is very large, $\Delta{y_{NL}}$ is positive (negative) for normal (anomalous) dispersion~\cite{piredda2007slow,https://doi.org/10.48550/arxiv.2301.13300,banerjee2022anomalous} with $n_g^{NL}>n_0^{-1}$ ($n_g^{NL}<n_0^{-1}$), respectively. While the nonlinearity experienced by a continuous wave (CW) beam propagating in a thermal medium can typically be approximated as instantaneous~\cite{marcucci2019optical}, the coupling between thermal nonlocality and medium motion introduces non-instantaneous effects, as delineated by the generalization of Eq. (4) to nonlocal nonlinear refraction in Eq. (29), elaborated upon in Sec. III. 
We will later also elaborate further on the effect of the speed of the medium on $n_g^{NL}$. The nonlinear photon drag effect can be tuned using the movement speed of the medium, creating a range of transverse shifts. To this point, the discussion has focused on the purely linear motion of an isotropic medium, so we next transition to rotation-based transverse drag.


\section{Rotation and anisotropy~\label{sec:2}}

\subsection{Media in rotation}

\begin{figure}[b!]
\begin{centering}
\includegraphics[width=0.8\columnwidth]{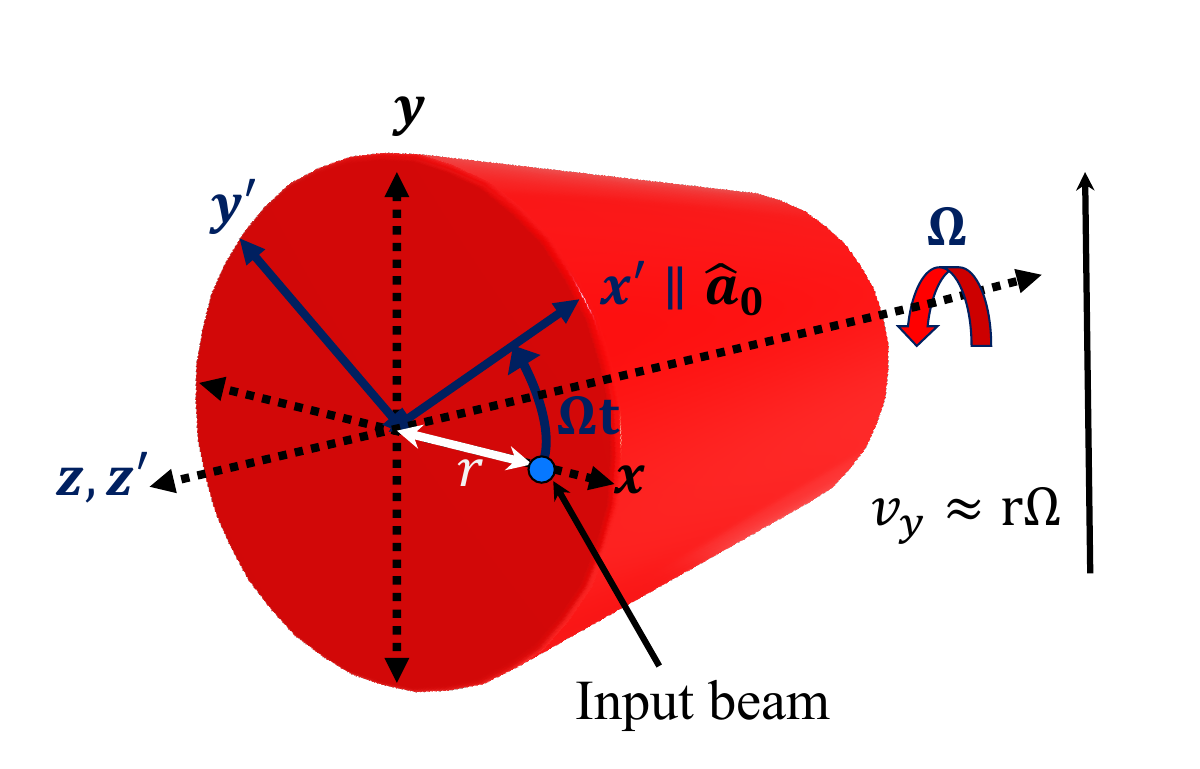}
\caption{{\bf A laser beam incident on a rotating medium far from the center of rotation.} Far from the center, the beam experiences the tangential component of the velocity in the direction according to the sign of the angular velocity, $\Omega$, that rotates about the z-axis. Two frames of reference are also shown. The lab frame is shown in ($x, y, z$) and the crystal frame is ($x', y', z'$).}
\label{fig:0a}
\end{centering}
\end{figure}

\begin{figure}[htp!]
\centering    
\includegraphics[width=0.69\columnwidth]{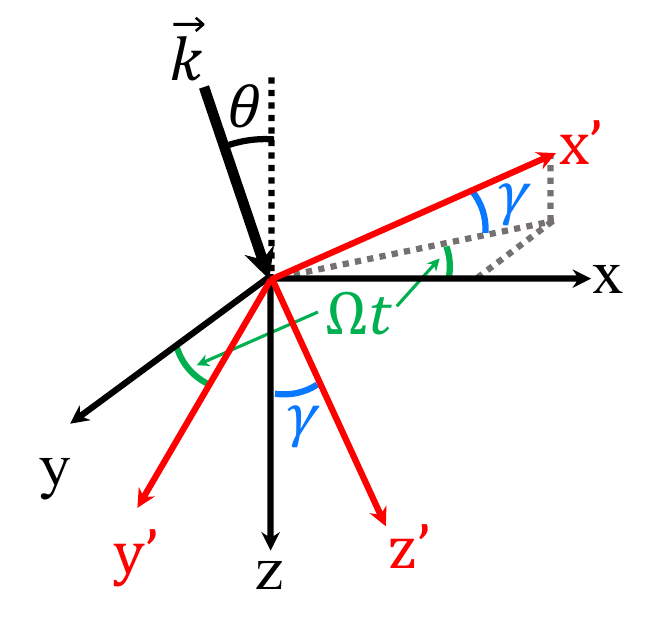}
\caption{\textbf{Two frames of reference.} The reference frames are laboratory (x,y,z) and crystal (x',y',z'). The wave vector comes in at an angle $\theta$ in the x-z plane.}
\label{fig:7}
\end{figure}

We can approximate rotation to fast translational motion by considering the tangential component of rotation for a beam far from the center of rotation ($r>w_0$) $r$ is the distance from the center of rotation to the center of the beam, and $w_0$ is the beam radius as seen in Fig.~\ref{fig:0a}. In the slow light regime where $|n_g^{NL}|\gg n_0$, and accounting for the rotation, the nonlinear photon drag effect becomes
\begin{equation}
\Delta{y_{NL}}\approx \frac{Lr\Omega}{c}n_{g}^{NL},
\label{eq:dragnonlinapprox2}
\end{equation}
where $\Omega$ is the medium rotation speed. 

The medium must easily rotate about the transverse plane and have a large group index to maximize the nonlinear photon drag effect. A suitable candidate would be a ruby rod~\cite{https://doi.org/10.48550/arxiv.2301.13300}, which has been used to investigate slow and fast light effects~\cite{bigelow2003superluminal,bigelow2003observation,lee1990measurements,cerdan2008simple,yang2005slow,wisniewski2014experimental,franke2011rotary}. Ruby also has measurable optical nonlinearities, including Kerr-type nonlinearity~\cite{lee1990measurements,catunda1986differential,kumari2014studies,boothroyd1989determination}, and in addition, it is an anisotropic medium that exhibits birefringence. We must account for the change in the group index but not the phase index since the group index can be much larger due to strong local dispersion effects. Hence, our model must incorporate the different refractive indices along its crystal structure's o- and e-axes. Motivated by this example, we will incorporate a rotating reference frame with a birefringent, nonlinear material into our model.

\subsection{Rotating reference frame with anisotropy\label{sec:eps}}

Consider a solid, birefringent ruby rod rotating about the z-axis with a constant rotation speed, as in Fig.~\ref{fig:0a}. We describe how incoming monochromatic light interacts with this medium using a vector wave equation for the electric field derived from Maxwell's equations
\begin{equation}
\mathbf{k}\times\left(\mathbf{k}\times\mathbf{E}\right)+\frac{\omega^2}{c^2}\mathbf{\epsilon}\mathbf{E}=\mathbf{0},
\label{eq:WEk}
\end{equation}
where $\mathbf{k}$ is the wave vector, $\omega$ is the frequency, $\epsilon$ is the dielectric tensor, and $\mathbf{E}$ is the electric field.

To properly describe the system, we must define the vector quantities in two reference frames: the lab frame $(x,y,z)$ and the rotating crystal frame $(x', y', z')$. In the lab frame, the orthonormal basis of unit vectors is
$\hat{\mathbf{x}}, \hat{\mathbf{y}}, \hat{\mathbf{z}}$. The crystal rotates with constant angular velocity $\mathbf{\Omega}$, and away from the center of rotation, the crystal moves with tangential velocity $\mathbf{v}=\mathbf{\Omega}\times\mathbf{r}$, where $\mathbf{r}=x\hat{\mathbf{x}}+y\hat{\mathbf{y}}+z\hat{\mathbf{z}}$. We write the crystal frame as another orthonormal basis $\hat{\mathbf{x}}', \hat{\mathbf{y}}', \hat{\mathbf{z}}'$, and take $\hat{\mathbf{x}}'=\hat{\mathbf{a}}_0$, where $\hat{\mathbf{a}}_0$ is the crystal optic axis~\cite{fowles1989introduction}. The crystal frame coordinates are accessed by applying a rotation matrix of an angle $-\Omega t$ about $z$-axis,

\begin{equation}
\mathbf{R_z}(-\Omega t) = \left(\begin{array}{ccc}\cos\left(\Omega t\right) & \sin\left(\Omega t\right) & 0\\-\sin\left(\Omega t\right) & \cos\left(\Omega t\right) & 0\\0 & 0 & 1\end{array}\right),
\end{equation}
to the lab frame basis vectors and vice versa. In the simplest case, the crystal basis is exactly aligned with the crystal axes, but generally, the system has a tilt angle, $\gamma$. We suppose that $\gamma$ rotates the x-z plane (i.e. about the y-axis). Now consider $\gamma\neq0$, as shown in Fig. \ref{fig:7}.

\subsubsection{Crystal imperfectly aligned, $\gamma\neq0$}

An angular difference between reference frames, $\gamma$, further induces a rotation considered to be between the optic axis, $\hat{\mathbf{a}}_0$, and the x-z plane, again setting $\hat{\mathbf{x}}' =\hat{\mathbf{a}}_0$ and $\hat{\mathbf{y}}', \hat{\mathbf{z}}'$ accordingly. The crystal frame becomes
\begin{equation}
\begin{array}{ccl}
x' &=& \cos(\gamma)\cos\left(\Omega t\right)x+\cos(\gamma)\sin\left(\Omega t\right)y+\sin(\gamma)z,\\
y' &=& -\sin\left(\Omega t\right)x+\cos\left(\Omega t\right)y,\\
z' &=& \sin(\gamma)\cos\left(\Omega t\right)x+\sin(\gamma)\sin\left(\Omega t\right)y+\cos(\gamma)z.
\end{array}
\label{eq:crystalgamma}
\end{equation}
Incorporating the tilt angle into the change of basis matrix $\textbf{C}$, we find 
\begin{equation}
\begin{aligned}
   \mathbf{C} &= \mathbf{R_y}\left(\gamma\right) \mathbf{R_z}\left(-\Omega t\right) \\&=
        \begin{pmatrix}
        \cos(\gamma)\cos\left(\Omega t\right) & \cos(\gamma)\sin\left(\Omega t\right) & \sin(\gamma)\\
        -\sin\left(\Omega t\right) & \cos\left(\Omega t\right) & 0\\
        -\sin(\gamma)\cos\left(\Omega t\right) &  -\sin(\gamma)\sin\left(\Omega t\right) & \cos(\gamma)
        \end{pmatrix}, 
\end{aligned}
\end{equation}
which comprises inverse rotation matrices about the z- and y-axes $\mathbf{R_z}(-\Omega t)$ and $\mathbf{R_y}\left(\gamma\right)$,  respectively, as seen by the crystal frame and shown in Fig. \ref{fig:7}. 

Incoming light in the crystal frame will see the ordinary ($n_o$) and extraordinary ($n_e$) refractive indices. We must consider the crystal symmetry when switching reference frames and its effect on the dielectric tensor. Returning to the lab frame, we apply $\mathbf{C}^{-1}$ to the dielectric tensor, neglect terms $O[(\delta n)^2]$, assuming $\delta n =n_o-n_e \ll1$ (e.g., $\delta n= -0.008$  for a uniaxial ruby rod), and find
\begin{equation}
\mathbf{\epsilon}\left(\gamma,\Omega t \right)=\mathbf{C}^{-1}\mathbf{\epsilon}\mathbf{C}=\mathbf{\epsilon}'+\mathbf{\epsilon}''\left(\gamma,\Omega t\right).
\label{eq:lab_eps_gamma}
\end{equation}
The optic axis is aligned to $z$ such that the above transformations allow transfer into the lab frame even when the axis of rotation ($z'$) is not perfectly aligned. Therefore, we take the optical axis perpendicular to the axis of rotation and find the dielectric tensor~\cite{fowles1989introduction} as
\begin{equation}
\epsilon'=\epsilon_0\left(\begin{array}{ccc}n_e^2 & 0 & 0\\0 & n_o^2 & 0\\0 & 0 & n_o^2\end{array}\right),
\label{eq:crystal_eps}
\end{equation}
\begin{widetext}
and $\epsilon''(\gamma,\Omega t)$ is
\begin{equation}
\begin{aligned}
      \epsilon''(\gamma,\Omega t)= 2 \epsilon_0 \delta n \cos^2(\gamma)&
        \begin{pmatrix}
        \sin^2(\Omega t) + \sec^2(\gamma) - 1  & -\sin(2\Omega t)/2  & \cos\left(\Omega t\right)\tan(\gamma)\\
        -\sin(2\Omega t)/2 & \sin^2(\Omega t) & \sin\left(\Omega t\right)\tan(\gamma)\\
        -\cos\left(\Omega t\right)\tan(\gamma) &  -\sin\left(\Omega t\right)\tan(\gamma) & \tan^2(\gamma) \end{pmatrix}\\\xrightarrow{\gamma=0}2\epsilon_0n_o\delta n&\left(\begin{array}{ccc}
\sin^2(\Omega t)&-\frac12\sin(2\Omega t)&0\\
-\frac12\sin(2\Omega t)&-\sin^2(\Omega t)&0\\
0&0&0
\end{array}\right).  
\end{aligned}
\end{equation}
Both the dielectric tensor and the interacting fields are needed to understand light propagation through the medium. So far, we have described the dielectric permittivity, including birefringence, tilt angle, and medium rotation. Next, we will address the fields.
\end{widetext}

\subsection{Propagating fields inside a rotating medium}
\subsubsection{Crystal perfectly aligned, $\gamma=0$}
Consider a monochromatic field propagating through a linear medium 
\begin{equation}
    \mathbf{E}=\mathbf{E}_0 e^{\imath\left(\mathbf{k}\cdot\mathbf{r}-\omega t\right)},
\end{equation}
under the assumption of weak birefringence ($\delta n\ll1$) and non-relativistic rotation speeds, $r\Omega\ll c$ (for which $\frac{\partial^2}{\partial t^2}\mathbf{\epsilon}\mathbf{E}\simeq\mathbf{\epsilon}\frac{\partial^2}{\partial t^2}\mathbf{E}$) ensuring that the time for the light to fully propagate through the medium is short compared to all other timescales.

In the lab frame, we solve the vector wave equation  Eq.~\eqref{eq:WEk} as a linear system of variables satisfying $\mathbf{A}\cdot\mathbf{E} = \mathbf{0}$, where $\mathbf{A}=\mathbf{k}^2-\frac{\omega^2}{c^2}\mathbf{\epsilon}$. We only find non-trivial solutions of $\mathbf{A}\cdot\mathbf{E}$ if the determinant of the coefficient matrix is non-null (i.e., $\mathbf{k}^2-\frac{\omega^2}{c^2}\mathbf{\epsilon}\neq0$). Using the dielectric tensor in Eq.~\eqref{eq:lab_eps_gamma} and the associated monochromatic field $\mathbf{E}$ for a rotating birefringent medium, we can solve $\mathbf{A}\cdot\mathbf{E}=0$. We must suppose that the initial wave vector $\mathbf{k}$ comes in at an angle $\theta$ between the optic axis and the z-axis, where at time $t=0$, $\hat{\mathbf{x}}=\hat{\mathbf{x}}' =\hat{\mathbf{a}}_0$ (See Fig.~\ref{fig:7}). Neglecting all the terms $O\left[(n_o^2-n_e^2)^2\right]$ in $\mathbf{A}\cdot\mathbf{E}$ and supposing that the crystal is perfectly aligned with the rotation axis ($\gamma=0$), then the conditions for which the wave vector coordinates $k_x, k_z$ ($k_y=0$), with $n_{e, i}=\sqrt{\epsilon_{ii}}$, to resolve non-trivial solutions in the lab frame are
\begin{equation}
\frac{k_x^2}{n_{e,2}^2}+\frac{k_z^2}{n_{e,2}^2}=\frac{\omega^2}{c^2},
\label{eq:circumference}
\end{equation}
\begin{equation}
\frac{k_x^2}{n_{e,3}^2}+\frac{k_z^2}{n_{e,1}^2}=\frac{\omega^2}{c^2},
\label{eq:ellipse}
\end{equation}
where 
\begin{equation}
    \begin{aligned}
      &n_{e,1}(\Omega t)=n_e+\delta n \sin^2(\Omega t),\\ &n_{e,2}(\Omega t)=n_o-\delta n \sin^2(\Omega t) \\ &n_{e,3}(\Omega t) \equiv n_o. 
    \end{aligned}
    \label{eq:ne1}
\end{equation}
The latter set of equations hold true only in the zero tilt-angle case. Equations for the non-zero tilt-angle case are reported in the Supplementary Materials.
\begin{figure}[!t]
\begin{center}
\includegraphics[width=0.7\columnwidth]{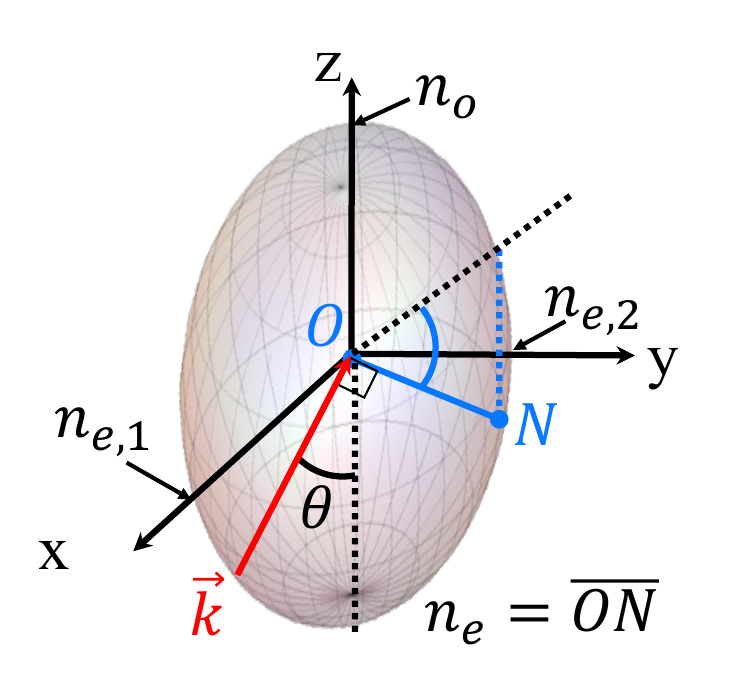}
\end{center}
\caption{
{\bf The refractive index ellipsoid in the lab frame}. Four refractive indices are labelled on the ellipsoid, $n_o$ as a projection onto the z-axis, $n_e$, which is equal to the length from O to N, defined as $\overline{ON}$, and $n_{e,1}$ and $n_{e,2}$ defined in Eq.~\eqref{eq:ne1}. $\vec{k}$ represents the wave vector of the incoming light, perpendicular to $\overline{ON}$.}
\label{fig:5}
\end{figure}
The quantities $n_{e,1}$ and $n_{e,2}$ represent the new refractive indices along $x$, and $y$, respectively. While in the crystal, the refractive indices ellipsoid reads as
\begin{equation}
\frac{(x')^2}{n_e^2}+\frac{(y')^2}{n_o^2}+\frac{(z')^2}{n_o^2}=1,
\label{eq:ellipsoid_cr}
\end{equation}
and in the lab frame, its expression is
\begin{equation}
\frac{x^2}{n_{e,1}^2}+\frac{y^2}{n_{e,2}^2}+\frac{z^2}{n_o^2}=1,
\label{eq:ellipsoid_lab}
\end{equation}
as sketched in Fig.~\ref{fig:5}.

Equation~\eqref{eq:ellipsoid_lab} demonstrates that, even if the crystal is uniaxial, the rotation in the lab frame results in a biaxial-like crystal with time-dependent birefringence. In a birefringent medium, two beams, the ordinary and extraordinary, are typically seen at the output face (see supplementary materials for more details). At certain conditions, a fixed input polarization state is sometimes purely ordinary or extraordinary, resulting in only one beam at the output.
This inherent birefringence, or lack thereof, causes the incoming linearly polarized light to interchange between o- and e-beams. Ultimately, predicting the exit position, angle, and relative intensity will be necessary to compare to experiments. More details on this are reported in the Supplementary Materials, including a modified Snell's law to understand the separation in output between o- and e-beams and the full derivation of the refractive indices in the case of the presence of the tilt angle, $\gamma$.

\subsubsection{Crystal imperfectly aligned, $\gamma\neq0$}
Assuming normal incidence and non-zero tilt-angle, we find the electric field for linearly polarized light interacting with the medium as
\begin{equation}
\begin{aligned}
\mathbf{E}(x,y,z,\Omega t,\gamma)&=\hat{\mathbf{E}}_o(\Omega t,\gamma)A(x,y,z)e^{\imath\left[\mathbf{k}_o(\Omega t,\gamma)\cdot\mathbf{r}-\omega t\right]}\\&+\hat{\mathbf{E}}_e(\Omega t,\gamma)B(x,y,z)e^{\imath\left[\mathbf{k}_e(\Omega t,\gamma)\cdot\mathbf{r}-\omega t\right]}.
\end{aligned}
\label{eq:efield}
\end{equation}
where $\hat{\mathbf{E}}_o = \hat{\mathbf{x}}'$, $\hat{\mathbf{E}}_e = \hat{\mathbf{y}}'$, and $A(x,y,z)$ and $B(x,y,z)$ are spatially varying functions defined later. The rotation of the linearly polarized light in a rotating birefringent medium leads to linear dynamics in the output transverse trajectories producing figure-eight-like patterns. The crystal is aligned such that, in the linear regime, the o-beam passes straight through, and the e-beam rotates around the o-beam at a rotation speed of $\Omega$. The behaviour of $\Omega$ on these trajectories is discussed later for both linear and nonlinear regimes. 

Because the output position of each of the birefringently separated beams moves rapidly while also varying in relative intensity, experiments can benefit from instead tracking the centre of intensity (COI) of the two beams~\cite{hogan2023beam}. Therefore, to facilitate comparisons to such experiments, we also model the COI. The COI can be constructed by first looking at the positions of two beams (o- and e-) separately, then taking the position of the average intensity of the two beams. For a rotating, birefringent medium, COI trajectories typically follow figure-eight-like patterns, with transverse displacement of the figure eight indicating the mean transverse displacement of the ordinary and extraordinary beams, while deformation of the figure-eight pattern indicates strong thermal nonlinearity leading to an index gradient, based on the results of ~\cite{hogan2023beam}.



Applying Eq.~\eqref{eq:WEk} to the newly found fields of Eq.~\eqref{eq:efield}, we again find non-trivial solutions to $\mathbf{A}\cdot\mathbf{E} = \mathbf{0}$ resulting in refractive indices in the crystal frame of the form (full derivation in Supplementary Materials)
\begin{equation}
\begin{aligned}
    n_{e,1}(\Omega t, \gamma)&\simeq n_e+\delta n \cos^2(\gamma)\left[\sin^2(\Omega t)+\tan^2(\gamma)\right],
    \\n_{e,2}(\Omega t, \gamma)&\simeq n_o-\delta n \cos^2(\gamma)\sin^2(\Omega t),
    \\n_{e,3}(\Omega t, \gamma)&\simeq n_o-\delta n\sin^2(\gamma).
    \label{eq:nvecs}
\end{aligned}
\end{equation}
In the limit of $\gamma=0$, one recovers Eq.~\eqref{eq:ne1}. 
We now model how the presence of the new refractive indices will alter the beams' propagation in the crystal. 

We model the propagation of light starting from the standard vector wave equation for the electric field derived from Maxwell's equations~\cite{marcucci2019optical} to obtain a generalized nonlinear Schrödinger equation (NLSE) for the o- and e-beam. We first substitute Eq.~\eqref{eq:efield} and the eigenvalue solutions, Eq.~\eqref{eq:nvecs} into the vector wave equation. We then assume $\nabla\cdot\mathbf{E}\simeq0$, as the dielectric tensor dependence on the spatial coordinates can be neglected in the limit $v\ll c$, which also ensures its negligible temporal derivative. We further assume weak birefringence, a slow-varying envelope and paraxial approximations,
\begin{equation}
\begin{aligned}
&\hat{\mathbf{E}}_o\nabla^2\left[Ae^{\imath\left(\mathbf{k}_o\cdot\mathbf{r}\right)}\right]+\hat{\mathbf{E}}_e\nabla^2\left[Be^{\imath\left(\mathbf{k}_e\cdot\mathbf{r}\right)}\right]\\
&+\frac{\omega_0^2}{\epsilon_0c^2}\left[Ae^{\imath\left(\mathbf{k}_o\cdot\mathbf{r}\right)}\mathbf{\epsilon}\hat{\mathbf{E}}_o+Be^{\imath\left(\mathbf{k}_e\cdot\mathbf{r}\right)}\mathbf{\epsilon}\hat{\mathbf{E}}_e\right]=\mathbf{0}.
\label{eq:WE2}
\end{aligned}
\end{equation}
and by making the substitutions
\begin{equation}
\begin{aligned}
 &A = a(x,y,z) \exp\Bigg({\imath \frac{\left(n_o^{\mathrm{eff}^2}-n_o^2\right)\omega_0^2}{2k_o^2c^2}\mathbf{\Delta k_o'}\cdot\hat{r}}\Bigg),\\& B = b(x,y,z) \exp\left({\imath\frac{\left(n_e^{\mathrm{eff}^2}-n_e^2\cos\left(\gamma\right)^2\right)\omega_0^2}{2k_e^2\cos\left(\gamma\right)^2c^2}\mathbf{\Delta k_e'}\cdot\hat{r}}\right),
\end{aligned}
\end{equation}
into Eq. \eqref{eq:WE2}, we can find two NLSEs that can be written out for the o- and e-beams. However, we must also introduce a nonlinear index gradient caused by nonlinear refraction, $\Delta n_{NL}(I)$, where $I$ is the beam intensity add to dielectric tensor in Eq.~\eqref{eq:lab_eps_gamma} (See Supplementary Materials for details). The thermal nonlinear response of $\Delta n_{NL}$ is important only when the timescales of interactions are long (i.e., non-instantaneous) and the nonlocal response is strong. With these substitutions, we arrive at two coupled nonlinear Schrödinger equations
\begin{widetext}
\begin{equation}
\begin{aligned}
 &\partial_z a=\frac{\imath}{2k_o}\nabla_{\perp}^2a-\frac{\imath k_o}{n_o}\Delta n_{NL}a- \frac{n_{g,o}^{NL}}{c}\partial_y a,
\\\partial_z b=\frac{\imath}{2k_e\cos^2(\gamma)}\nabla_{\perp}^2b&+\frac{\imath k_e}{n_e\cos^2(\gamma)}\Delta n_{NL}b+2 \tan(\gamma)\left(\cos(\Omega t)\partial_x b +\sin(\Omega t)\partial_y b\right)-\frac{n_{g,e}^{NL}}{c}\partial_y b.  
\end{aligned}
\label{eq:WE_nga}
\end{equation}
\end{widetext}
where 
\begin{equation}
    n_{g,o}^{NL}=\frac{\omega_0^2 \left(n_o^{\mathrm{eff}^2}-n_o^2\right)\Delta k}{2k_o^3c^2},
\end{equation}
and
\begin{equation}
    n_{g,e}^{NL}=\frac{\omega_0^2 \left(n_e^{\mathrm{eff}^2}-n_e^2\cos^2(\gamma)\right)\Delta k}{2k_e^3\cos^4(\gamma)c^2}.
\end{equation}

Simulating these two NLSEs, we can extract the output COI transverse trajectories and the amount of transverse shift. Simulations also showed that these effects of $\Delta n_{NL} $ are small (See Supplementary materials) in comparison to the nonlinear response of the group index, largely due to the interaction timescales being much shorter for the group index. $\Delta n_{NL}$ is described as a first-order nonlinear correction in the crystal frame to the dielectric permittivity and is dependent on rotation speed and tilt angle. 
In thermal media~\cite{marcucci2019optical}, 
\begin{equation}
\Delta n_{NL}' =\left(\frac{\partial n}{\partial T}\right)_0 \Delta T(\mathbf{r}'),
\label{eq:n_T}
\end{equation}
where $\left(\frac{\partial n}{\partial T}\right)_0$ is the medium's thermo-optic coefficient at thermal equilibrium (steady-state response) and $\Delta T(\mathbf{r}')$ is the temperature variation about the point $\mathbf{r}'=(x',y',z')$. $\Delta T(\mathbf{r}'$) for a stationary medium is governed by the $3$D heat equation
\begin{equation}
\left(\partial^2_{x'}+\partial^2_{y'}+\partial^2_{z'}\right)\Delta T(\mathbf{r}')=-\gamma|\mathbf{E}'(\mathbf{r}')|^2,
\label{eq:temperature0}
\end{equation}
with $\gamma_l=(L_{loss}\rho_0c_P D_T)^{-1}$, where $L_{loss}$ is the loss characteristic length, $\rho_0$ the material density, $c_P$ the specific heat at constant pressure, and $D_T$ is the thermal diffusivity.  Assuming absorption is low ($L\ll L_{loss}$), we find $\Delta T(\mathbf{r}') \sim \Delta T_{\perp}(\mathbf{r_{\perp}}')$ and $\partial_{z'} I'(\mathbf{r}')\sim 0$, and lastly that
\begin{equation}
\Delta n_{NL}(\Omega t,\gamma)= n_2 \iint \mathrm{d}\widetilde{x} \mathrm{d}\widetilde{y} K_\gamma \left(\Delta x, \Delta y, \Omega t \right) I(\widetilde{x},\widetilde{y}) - n_{o,e}, 
\label{eq:nnl}
\end{equation}
where $K_\gamma$ is the nonlinear nonlocal kernel function affected by the weak birefringence, written as
\begin{equation}
 K_\gamma \left(\Omega t, \gamma\right)= K' 
    \left\{ \begin{array}{ll}
        \cos(\gamma)\left(\cos(\Omega t) x+\sin(\Omega t) y\right)\\
 -\sin(\Omega t) x+ \cos(\Omega t) y \\
    \end{array} \right\}.
\end{equation}

Although $\Delta n_{NL}$ can be large at times, we consider it as small and focus on the discussion of the effects of the nonlinear group index. 

\vspace{-5mm}
\section{Influence of the $n_g^{NL}$ \label{sec:5}}
Starting from the general definition, we have the group refractive indices for the o- and e-beams as 
\begin{equation}
\begin{aligned}
    n_g^{o,e}&=n_{o,e}+\left(\omega \frac{\partial n_{o,e}}{\partial \omega}\right)_{\omega_o}.
\end{aligned}
\end{equation}

We can represent the individual dependencies on the rotation speed and input intensities of the group indices for the o- and e-beams by utilizing the Lorentz transformation, $\Delta k =-\frac{\omega_0\Omega r}{c^2}$, to find
\begin{equation}
\begin{aligned}
    &n_g^{o}=n_{o}+\frac{\left(n_o^{\mathrm{eff}^2}-n_{o}^2+n_o^g n_2^g I_{a}\right)}{2n_{o}^3}\frac{\Omega x_0}{c},\\&    n_g^{e}=n_{e}+\frac{\left(n_e^{\mathrm{eff}^2}-n_{e}^2+n_e^g n_2^g I_{b}\right)}{2n_{e}^3\cos^4(\gamma)}\frac{\Omega x_0}{c},
\end{aligned}
\end{equation}
where 
\begin{equation}
\begin{aligned}
&I_{a}\simeq |(-h) \sin(\Omega t) +v \cos(\Omega t)|^2 I_0,\\&I_{b}\simeq |h\cos(\Omega t)+ v\sin(\Omega t)|^2 I_0.
\end{aligned}
\end{equation}
These intensities, $I_a$ and $I_b,$ are the individual intensities of the o- and e-beams, respectively. The variables $h$ and $v$ represent the input polarization in the lab frame, whether  \textit{H}- and \textit{V}-linear polarization. We have corrected the lowest order to the ordinary and extraordinary refractive indices that $n_o^{\mathrm{eff}}\approx n_g^o+\frac{1}{2}n_2^g I_a$
and $n_e^{\mathrm{eff}}\approx n_g^e+\frac{1}{2} n_2^g I_b$.
We can define a collective $n_g^{NL}$ that describes the COI of these two beams as 
\begin{equation}
    n_g^{NL}=n_g^0+n_2^g I,
\end{equation}
where $I=I_a+I_b$. Note that we approximate $n_g^0=(n_g^o+n_g^e)/2$ due to weak birefringence, and $n_2^gI\gg n_g^0$.

Since the $n_g^{NL}$ depends on both the optical and thermal nonlinear responses, one can write the full $n_g^{NL}$ as
\begin{equation}
n_g^{NL}= n_g^0+n_{2,\mathrm{opt}}^g I_0+n_{2,\mathrm{therm}}^g I_0.
\label{equation5a}
\end{equation}
and can be written as
\begin{equation}
n_g^{NL}= n_g^0+n_2^g I_0 \left(\alpha_s e^{-(\Omega-\Omega_0)/\Omega_s}-\alpha_f e^{-(\Omega-\Omega_0)/\Omega_f}\right).
\label{equation5b}
\end{equation}
Eq.~\eqref{equation5b} is a simplified, compact form of $n_g^{NL}$. These values are found by comparing them to our experiment~\cite{hogan2023beam}. The behaviour of $n_g^{NL}$ is modeled by a piecewise function about a characteristic speed $\Omega_c$, which captures the dynamics above and below $\Omega_c$. We use the piece-wise form for $n_g^{NL}$ in our simulations written as
\begin{equation}
n_g^{NL} =n_2^g I_0\times
    \left\{ \begin{array}{ll}
        (n_g^0/n_2^g I_0)-\alpha_f\exp\left(-\frac{\Omega-\Omega_c}{\Omega_f}\right) & \Omega\leq \Omega_c\\
        (n_g^0/n_2^g I_0)+\alpha_s\exp\left(-\frac{\Omega-\Omega_c}{\Omega_s}\right) & \Omega\geq \Omega_c.\\
    \end{array} \right.
    \label{equation3}
\end{equation}
The two pieces of the function are each individually valid in the asymptotic limit of extremely fast and slow rotation. With this form, we aim to match the results~\cite{hogan2023beam} by setting $n_2^g I_0=0.11\times10^7 $m$^2$/W, and fit constants $\alpha_s$ and $\alpha_f$ are taken to be 0.97 and 0.94, respectively. In \cite{hogan2023beam}, transverse beam shifts and transverse output beam trajectories were measured to quantify the deflection due to photon drag and other nonlinear effects and their effect on beam propagation. The tilt angle used is $\gamma=\pi/1800$. Other fit values for the piece-wise function are summarized in Table \ref{table2}. Offsets are described by $(n_g^0/n_2^g I_0)$, amplitudes $\alpha_{f}$, and $\alpha_{s}$, characteristic speeds $\Omega_{f}$ and $\Omega_{s}$ for thermal and optical nonlinear responses, respectively. All constants are strictly positive and retrieved for low-to-mid (5$\sim $100 deg/s) and mid-to-high (100$\sim $9000 deg/s) rotation speeds.

\begin{table*}
\centering
\begin{tabular}{|p{3cm}|p{3cm}|p{1.45cm}|p{0.95cm}|p{0.95cm}|p{0.95cm}|p{0.95cm}|p{0.95cm}|p{1.9cm}|}
\hline
{Intensity (W/cm$^2$)} & {Speed Range (deg/s)} &  \multicolumn{3}{c|}{{Variable}} & \multicolumn{3}{c|}{Value}\\
\hline\hline
{$6.4\times10^4$} & $5\sim100$ & $(n_g^0/n_2^g I_0)$ & $\alpha_f$ & $\Omega_f$ & 166 & 1.4 & 21\\
\hline 
{$6.4\times10^4$} & $100\sim9000$ & $(n_g^0/n_2^g I_0)$ & $\alpha_s$ & $\Omega_s$ & 11 & 151 & 576\\
\hline 
{$3.3\times10^5$} & $5\sim100$ & $(n_g^0/n_2^g I_0)$ & $\alpha_f$ & $\Omega_f$ & 617 & 80 & 49\\
\hline 
{$3.3\times10^5$} & $100\sim9000$ & $(n_g^0/n_2^g I_0)$ & $\alpha_s$ & $\Omega_s$ & 51 & 486 & 1190\\
\hline
\end{tabular}
\label{table:nonlin}
\caption{\textbf{Results of the phenomenological fit for $n_g^{NL}$.} Fit variables for the expressions in Eq.~\eqref{equation3} in the nonlinear ($P=100$ mW, $I=3.3\times10^4$ W/cm$^2$) and highly nonlinear ($P=520$ mW, $I=6.4\times10^5$ W/cm$^2$) regimes for low-to-mid speeds ($\Omega = 5\sim100$ deg/s) and mid-to-high speeds ($\Omega = 100\sim 9000$ deg/s) are shown.}
\label{table2}
\end{table*}

With complete knowledge of the system and the dynamics of $n_g^{NL}$, nonlinear propagation of the two coupled NLSEs is simulated using the Split-Step Fourier Method (SSFM) to extract the amount of transverse shift as well as the transverse trajectories at the crystal output. The results of the simulated NLSEs are discussed in the following section. Furthermore, the details of how the simulations are performed are described in the supplementary materials.

\section{Results and Discussion\label{sec:6}}

\begin{figure}[h!]
\begin{center}
\includegraphics[width=\columnwidth]{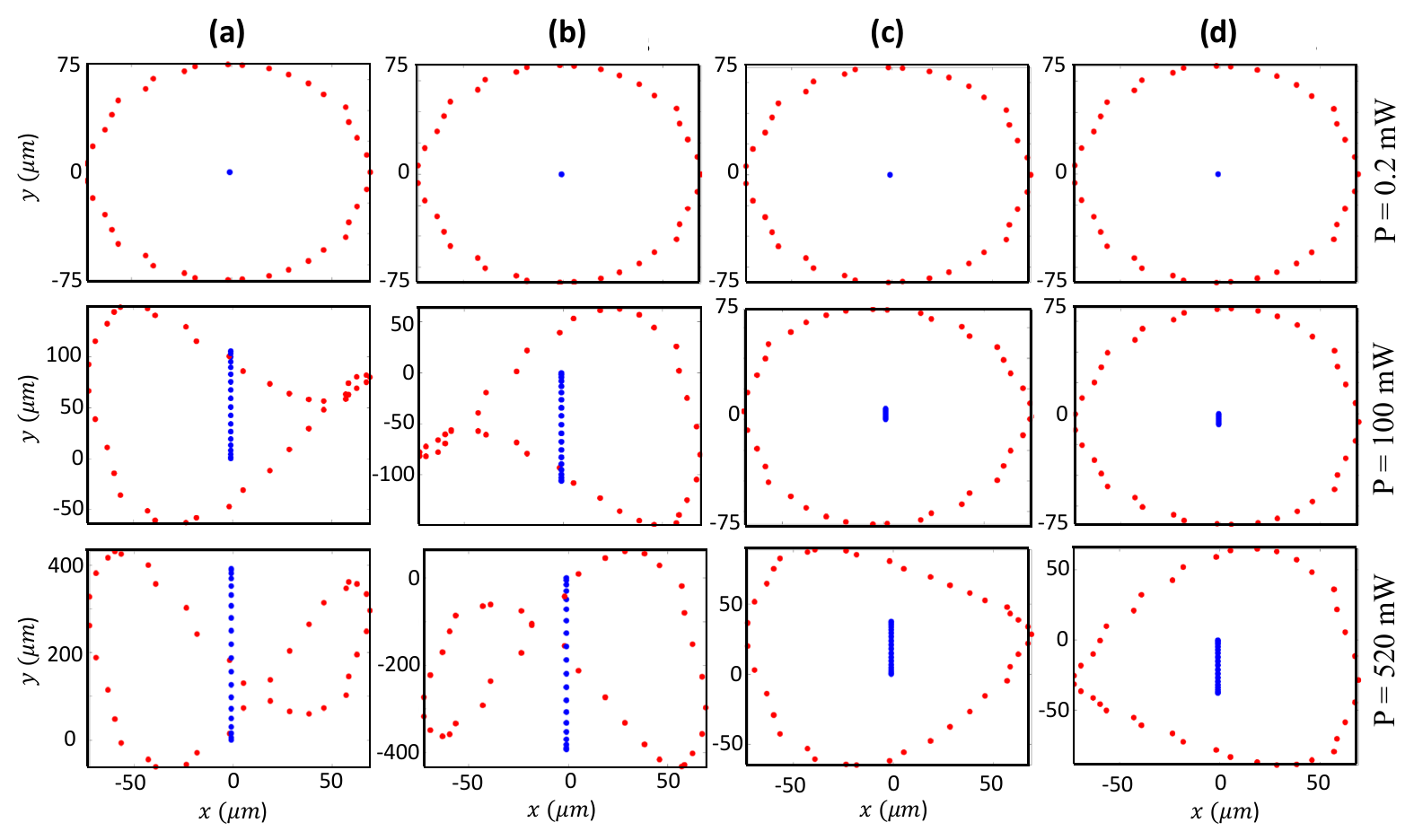}
\end{center}
\caption{\textbf{Transverse trajectories of the o- (blue) and e- (red) beam for three input powers at positive/negative low/high rotation speeds.}
Linear regime shows no transverse shift drag in either beam for different speeds (\textbf{(a)} $\Omega = -100 $ degs/s, \textbf{(b)} $\Omega = 100 $ degs/s, \textbf{(c)} $\Omega = -9000 $ degs/s, and \textbf{(d)} $\Omega = 9000 $ degs/s), while nonlinear regimes show increasing shift for a given speed. The magnitude of the shift is seen more clearly in the o-beam movement. At the same time, the e-beam shows deviations from a circular trajectory. The transverse shifts are also experienced by the e-beam, but since it is rotating, the local position changes, and the beam feels a different index gradient at each point.}
\label{fig:11a}
\end{figure}

\begin{figure}[t!]
\begin{center}
\includegraphics[width=\columnwidth]{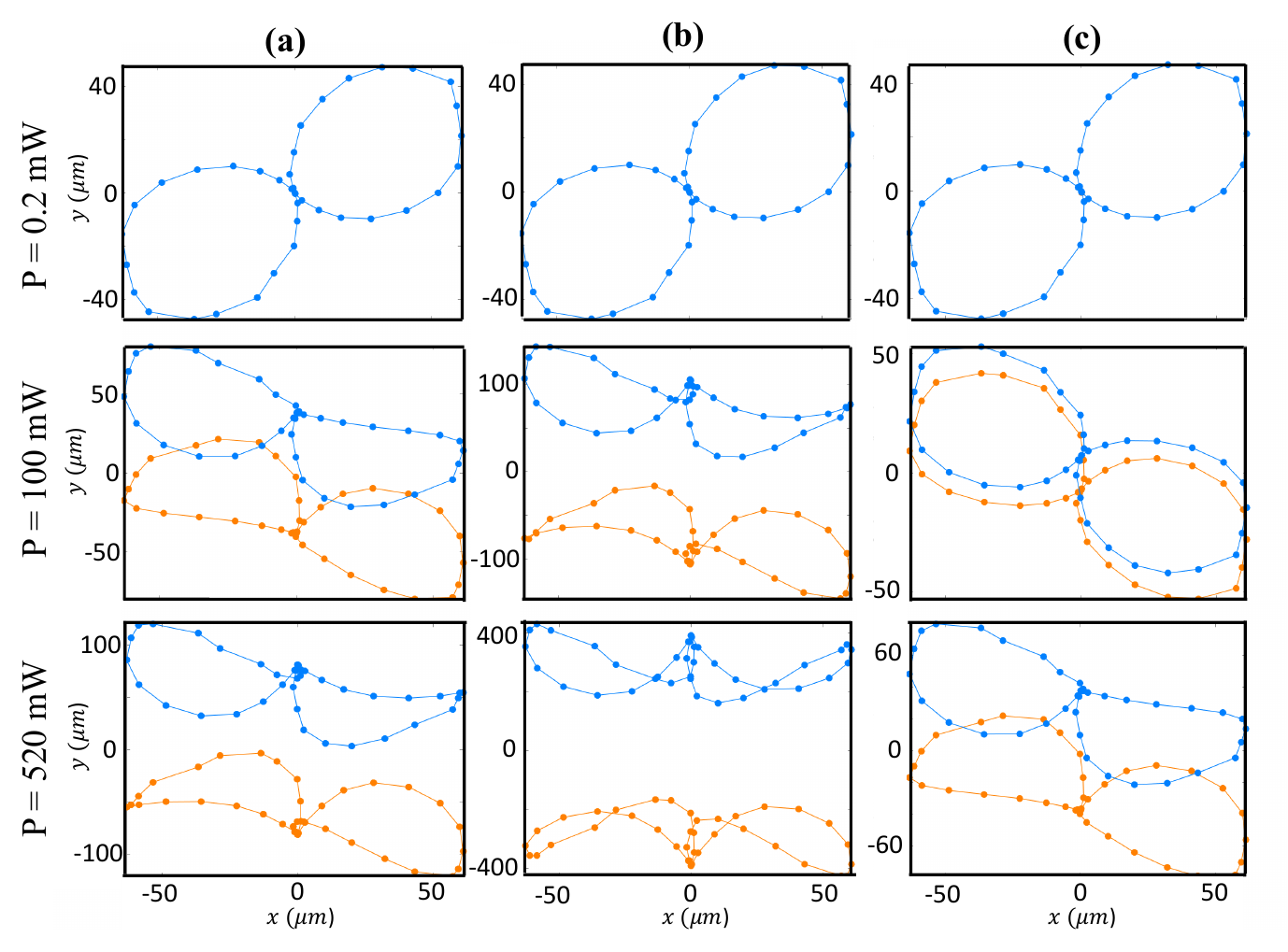}
\end{center}
\caption{\textbf{COI trajectories at the output of a 2-cm long ruby crystal for three input powers and three input speeds (a) $\Omega = \pm10$ deg/s, (b) $\Omega = \pm100$ deg/s, and (c) $\Omega = \pm9000$ deg/s.}  Simulated curves for the linear ($P = 0.2$ mW) regime show a figure-eight-like trajectory for the COI. In contrast, nonlinear ($P = 100$ mW) and highly nonlinear ($P = 520$ mW) regimes show deviations from a figure-eight and lateral displacement along $y$ (the direction of motion of the crystal). Blue and orange curves correspond to positive and negative rotation speeds, respectively.}
\label{fig:12}
\end{figure}

We simulate a 2-cm long ruby crystal illuminated with linearly polarized light for three input intensities, $I=1.3\times10^2$ ($P=200$ $\mu$W), $I=6.4\times10^4$ ($P=100$ mW), and $I=3.3\times10^5$ W/cm$^2$ ($P=520$ mW), over a range of rotation speeds $\Omega =$ 1 degs/s to 9000 degs/s to extract the amount of transverse shift and the transverse trajectories at the crystal output. 

\begin{figure*}
\begin{center}
\includegraphics[width=\textwidth]{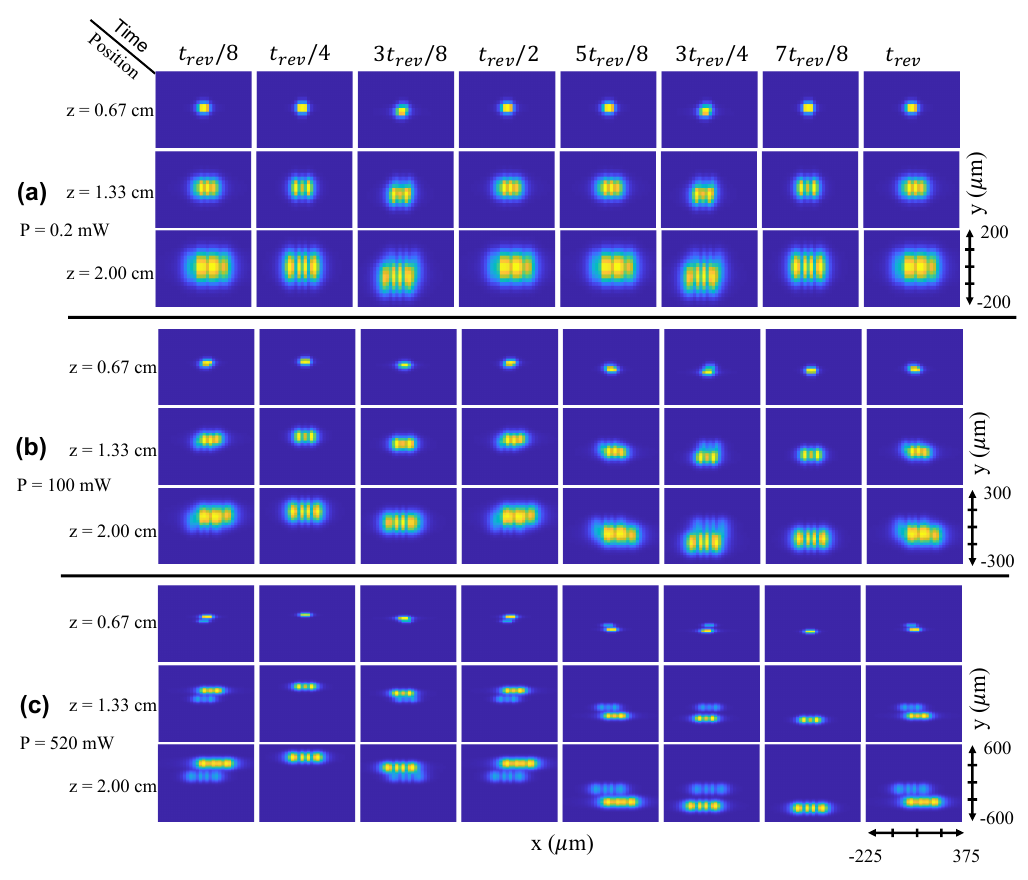}
\end{center}
\caption{\textbf{Simulated propagation and rotation effects on the E-field squared distribution.} The total E-field squared of the o- and e-beams is plotted for three z-positions within the ruby rod, $z=L/3$ cm, $z=2L/3$ cm, and $z=L$, where $L=2$ cm, and over eight different frames along the rotation in time, where $t_{rev}$ represents one full rotation, for a rotation speed of $\Omega=100$ degs/s for three regimes: \textbf{(a)} Linear ($P=0.2$ mW), \textbf{(b)} Nonlinear ($P=100$ mW) and \textbf{(c)} Highly Nonlinear ($P=520$ mW). Nonlinear effects can be observed, leaving imprinted traces of a beam from the index gradient when the input power is sufficiently high $P>100$ mW. The field trajectories widen with increasing power due to the nonlinear deflection due to the nonlinear group index.}
\label{fig:13}
\end{figure*}

Birefringence results in two distinct beams that propagate through the medium when linear polarized light is used, so the movement of both beams must be tracked. Figure~\ref{fig:11a} shows the trajectories of the o-beam in blue and the e-beam in red for three input powers in the low and high-speed regimes for positive and negative rotation speeds. The o-beam shows a transverse shift with increasing intensity, while the e-beam revolving around starts to cross and form a twisted pattern in the nonlinear regimes. From Eq.~\eqref{eq:WE_nga}, we see that the nonlinear group index works on the y-derivative of the o- and e-beams, so the rotation of the e-beam will see local variation in intensity while also following rotation. As such, the circular trajectory will distort. The distortions arise from the contributions of both optical and thermal nonlinear response, where slower rotation speeds distort based on thermal timescales and faster speeds with optical timescales. Moreover, since the nonlinear group index is intensity-dependent, these distortions are more dramatically seen for higher intensities. The optical and thermal effects are both present via $n_g^{NL}$, but contribute to a different extent based on whether the rotation speed is above or below a characteristic speed $\Omega_0$. The effects of distortions produce a negative shift if the nonlinear contribution of the group index is negative and larger than the linear group index. A negative shift would take place when the thermal nonlinear contribution is significantly larger than the nonlinear optical contribution, necessitating large intensities and a medium with a high damage threshold. 

Due to the presence of two beams, their COI results in transverse trajectories that create figure-eight-like patterns. We can see in Fig.~\ref{fig:12} that the linear dynamics of the COI trace out a figure eight. The twisting of the e-beam trajectories seen in Fig.~\ref{fig:11a} creates the twisted patterns seen in the nonlinear regimes of Fig.~\ref{fig:12}. The transverse shift and twisted patterns result from the thermal and optical nonlinear response the crystal impinges on the light as it passes through the crystal, resulting in distorted patterns for the COI. The trajectory patterns get distorted and transversely shifted relative to one another based on the nonlinear photon drag effect. Incorporating the birefringence, dispersion, and nonlinear response, we have observed that simulations produce transverse trajectories that match well with previous experiments~\cite{hogan2023beam}. The trajectories capture the linear and nonlinear dynamics of the system with good agreement, as seen in Fig. \ref{fig:12}. The traced-out COI trajectories for three rotation speeds of $\Omega =$ 10, 100, and 9000 degs/s correspond to i), ii), and iii), respectively. 

Although the central positions of the o- and e-beams can be tracked, the full electric field distribution should also be considered. Figure \ref{fig:13} shows the evolution of the square of the fields along z (top to bottom) and in time (left to right) for three input powers $P=0.2$ mW, $P=100$ mW, and $P=520$ mW at a rotation speed of $\Omega=100$ deg/s. At this rotation speed, both optical and thermal nonlinear responses are present. We examine the overlapped o- and e-fields along z at three positions: $z=L/3$, $z=2L/3$, and $z=L$, where $L=2$ cm. The linear ($P=0.2$ mW) dynamics are shown in Fig. \ref{fig:13}\textbf{(a)}, where beam size increases along z due to diffraction and rotation in time. Figures \ref{fig:13}\textbf{(b)} and \textbf{(c)} show imprinted beam traces due to $n_g^{NL}$ creating an index gradient impacting beam movement. At $\Omega=100$ degs/s, the observed effect from $n_g^{NL}$ is both optical and thermal nonlinear response. Therefore, any previous position of the beam is seen in nonlinear regimes for a given instant in time. No trace beam is seen for low intensity. The misshapen structure is a result of the overlapped beams, and so the beams at $z=L$ resemble the typical output of a crystal with sufficient propagation. 

\begin{figure}[t!]
\begin{center}
\includegraphics[width=0.95\columnwidth]{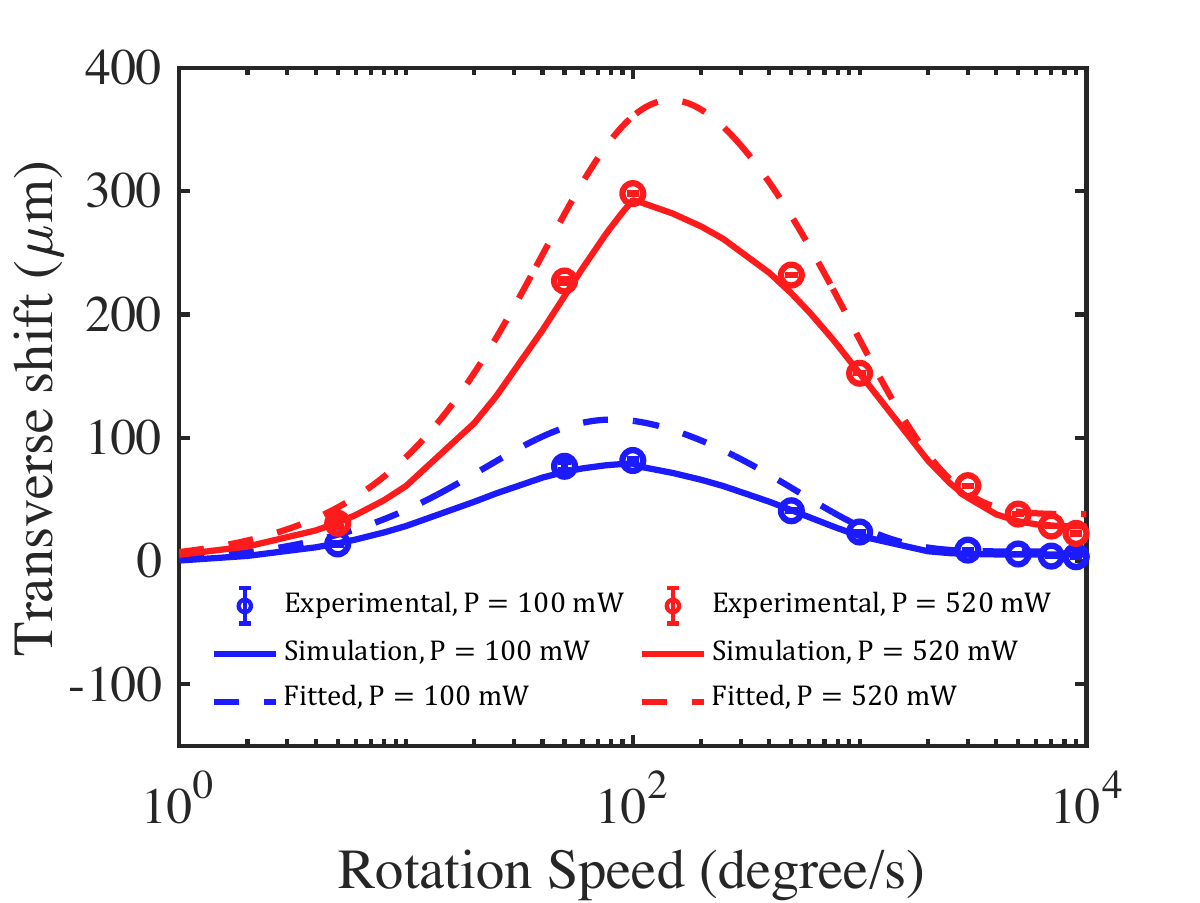}
\end{center}
\caption{\textbf{The transverse shift in the nonlinear and highly nonlinear regime for various rotation speeds.} The distribution shows a log-normal-like distribution about $\Omega=100$ degs/s. The phenomenological fit (dashed lines) suggests a peak closer to $\Omega=150$ degs/s, while the simulations and experimentally measured values suggest $\Omega=100$ degs/s. The fit overestimates the amount of shift in the transient regime around $\Omega=150$ degs/s as it is comprised of two exponentials. As a result, the behaviour is better represented in simulated data, which uses the two exponential behaviours separately as a piecewise function in Eq.~\eqref{equation3}, capturing the transient regime between optical and thermal contributions.}
\label{fig:9}
\end{figure}

Moreover, there is a significant increase in transverse movement with increasing power, as expected with the increased nonlinear deflection. Slower rotation speeds coincide with the thermal nonlinear response, while high speeds apply to optical effects that manipulate the fields over approximate integer multiples of one full rotation. The timescale associated with a full rotation is in the order of milliseconds, which is closer to the timescale of the optical response. The transverse shift can be extracted by looking at the average position of these trajectories, and the deflection dynamics can be more directly investigated.

Trajectories experience different amounts of transverse shifts in positive or negative directions for positive or negative speeds. Extracting for different rotation speeds and input powers, we plot the simulated transverse shifts against experimentally measured data points and phenomenological fits (dashed lines) in Fig. \ref{fig:9}. The close agreement suggests that our model fully describes the relevant linear and nonlinear processes producing transverse photon drag through a rapidly rotation ruby rod under slow light conditions. The phenomenological fit described by the continuous function in Eq.~\eqref{equation5b} suggests the characteristic rotation speed as $\Omega_c=150$ degs/s. However, we incorporate a piece-wise function into the NLSE to better describe the dynamics in the respective rotation speed ranges, resulting in the solid curves showing excellent agreement with experimental data. The fit reaches much larger transverse shifts around the transient regime of equal contributions of thermal and optical nonlinear response. The exponentials in Eq.~\eqref{equation5b} tend to larger values, causing larger than expected transverse shifts when combined and fit collectively. As a result, the behaviour is better represented using the two exponential behaviours separately as a piecewise function in Eq.~\eqref{equation3} around the transient regime.

\section{Conclusions\label{sec:8}}

We have developed a general theoretical model for the nonlinear propagation of light in moving media, particularly focusing on a rotating birefringent medium. Our theoretical model extends linear photon drag theory to the nonlinear regime by means of a nonlinear group index. A set of coupled generalized nonlinear Schrödinger equations was used to model the propagation of ordinary (o-) and extraordinary (e-) beams through the rotating nonlocal nonlinear medium. The coupled equations included a nonlinear group index, birefringence, rotating reference frames, and nonlinear optical and thermal responses that match the nonlinear dynamics of the system. The model also converges toward the expected trajectory shape of the linear dynamics but laterally shifts within the limits of fast and slow rotation speeds. The model was used to produce and study o- and e-beam trajectories, centre of intensity trajectories, electric field evolution and the amount of transverse shift at the crystal output~\cite{hogan2023beam}.

Our model accurately describes the dispersion and nonlinear response of beams propagating through a rotating nonlinear medium, allowing for predictable propagation directions and tunable control of the transverse shift at the output of the crystal. The tunable control of transverse position using power and rotation speed could be applied to beam-steering and sorting applications. Going forward, one could study the effects of input polarization on the transverse shift and beam distortion of different beam structures and beams carrying orbital angular momentum. One could also look at balancing dispersion and nonlinear effects to create solitons that hold their shape in a moving dispersive medium, suitable for classical communication or propagating non-distorted images.

\section*{Funding}
{R.W.B. and R.H. acknowledge support through the Natural Sciences and Engineering Research Council of Canada (NSERC) under Discovery Grant RGPIN/2017-06880, the Canada Research Chairs program under award 950-231657, and the Canada First Research Excellence Fund on Transformative Quantum Technologies under award 072623.  In addition, R.W.B. acknowledges support through the US Office of Naval Research under award N00014-19-1-2247 and MURI award N00014-20-1-2558. R.H. acknowledges support through Indigenous Affairs at the North Shore Micmac District Council.} A.S. acknowledges the support of NSERC under Grant PDF-546105-2020. B.B. also acknowledges the support of the Banting postdoctoral fellowship of NSERC.


\section*{Acknowledgements}
The authors would like to thank Xiaoqin Gao for her valuable advice in figure generation and structural formatting and Prof. Claudio Conti for insightful discussions on the non-instantaneity of thermal nonlinearity in moving media.

\section*{Disclosures}
We are unaware of any conflicts of interest associated with this work.
 
\bibliography{Main/References}


\newpage
\clearpage
\section*{Supplementary Materials}

This article focuses on creating a model to predict the beam path trajectory and evolution through a rapidly rotating, birefringent medium that can experience Kerr and thermal nonlinearities. The accompanying supplementary materials show derivations in length from those seen in the main text. For example, non-normal incident beam angle, which is often the case in experiments. Furthermore, the angle dependence is affected by the tilt angle relative to the axis of rotation and index gradients. All of these components will modify the propagation and transverse trajectories, as well as the transverse shifts.

\subsection*{Effects of misalignment on light propagation}
Another key aspect is incorporating a modified Snell's law to treat the birefringence and understand the beam separation with rotation. Looking at air-crystal and crystal-air interfaces, assuming that the light illuminates at some angle of incidence $\theta$, the new ordinary and extraordinary refractive indices in the lab frame are
\begin{equation}
\begin{aligned}
n_o(\Omega t)=n_{e,2}(\Omega t),
\end{aligned}
\label{eq:new_ord}
\end{equation} 
and
\begin{equation}
n_e(\Omega t,\theta)=\left[\frac{\cos^2(\theta)}{n_{e,1}^2(\Omega t)}+\frac{\sin^2(\theta)}{n_{e,3}^2(\Omega t)}\right]^{-\frac12},
\label{eq:new_ext}
\end{equation}
where $\Omega$ is the rotation speed, and $t$, time.

Once the light enters the birefringent material, two angles $\alpha$ and $\beta$ emerge within the crystal. A small deviation in angle can change the set of refractive indices and, therefore, the transverse trajectories at the output. Using standard algebra and modified Snell's law~\cite{fowles1989introduction},
\begin{equation}
\alpha(\Omega t,\theta)=\arcsin\left(\frac{n_{\mathrm{air}}}{n_{o}(\Omega t)}\sin(\theta)\right),
\label{eq:phi}
\end{equation}
\begin{equation}
\beta(\Omega t,\theta)=\arcsin\left(\frac{n_{\mathrm{air}}}{n_{e}(\Omega t,\theta)}\sin(\theta)\right)-\alpha(\Omega t,\theta).
\label{eq:delta}
\end{equation}

Upon propagation, the o-beam shifts transversely by a distance $d'$, found as the tangent of the angle $\alpha$ multiplied by the crystal length $L$
\begin{equation}
d'(\Omega t,\theta)=L\tan\left[\alpha(\Omega t,\theta)\right],
\label{eq:dprime}
\end{equation}

Similarly, we can find the distance between o- and e-beams $d$ 
\begin{equation}
d(\Omega t,\theta)=L\tan\left[\alpha(\Omega t,\theta)+\beta(\Omega t,\theta)\right]-d',
\label{eq:d}
\end{equation}
defined as the distance between the o- and e-beam in the x-y plane.
It is important to know the distance between the two beams since their separation can change as a result of nonlinear interactions when the intensity becomes large.

We can also describe the respective fields and how they will propagate within the crystal measuring in the lab frame. The sum of two fields describes the full field, the o- and e-fields, with spatially varying functions $A(x,y,z)$
and $B(x,y,z)$ as 
\begin{equation}
\begin{aligned}
\mathbf{E}(x,y,z,t,\Omega)&=\hat{\mathbf{E}}_o(\Omega t,\theta)A(x,y,z)e^{\imath\left[\mathbf{k}_o(\Omega t,\theta)\cdot\mathbf{r}-\omega t\right]}\\&+\hat{\mathbf{E}}_e(\Omega t,\theta)B(x,y,z)e^{\imath\left[\mathbf{k}_e(\Omega t,\theta)\cdot\mathbf{r}-\omega t\right]},
\label{eq:E}
\end{aligned}
\end{equation}
where $\hat{\mathbf{E}}_o = \hat{\mathbf{x}}'$, $\hat{\mathbf{E}}_e = \hat{\mathbf{y}}'$, and  $\hat{\mathbf{x}}'$, $\hat{\mathbf{y}}'$ are defined in the main text.

Normal incidence is often considered, but some crystals are non-optimally cut with respect to the easiest axis of rotation. Therefore, due to the crystal-cut optical axes, we consider the beam at normal incidence but imperfectly aligned ($\gamma\neq0$). 



Assuming a non-zero tilt-angle (the zero tilt-angle case can be obtained by imposing $\gamma=0$), we find the electric field for linearly polarized light interacting with the medium as
\begin{equation}
\begin{aligned}
\mathbf{E}(x,y,z,\Omega t,\gamma)&=\hat{\mathbf{E}}_o(\Omega t,\gamma)A(x,y,z)e^{\imath\left[\mathbf{k}_o(\Omega t,\gamma)\cdot\mathbf{r}-\omega t\right]}\\&+\hat{\mathbf{E}}_e(\Omega t,\gamma)B(x,y,z)e^{\imath\left[\mathbf{k}_e(\Omega t,\gamma)\cdot\mathbf{r}-\omega t\right]}.
\end{aligned}
\label{eq:efield}
\end{equation}
where polarizations and wave vectors now depend on both rotation and tilt angle. The crystal is aligned such that, in the linear regime, the o-beam passes straight through, and the e-beam rotates around the o-beam at a rate of $\Omega$. 


Using the fields of Eq.~\eqref{eq:efield}, non-trivial solutions of the system $\mathbf{A}\cdot\mathbf{E} = \mathbf{0}$ result from the condition $\det(\mathbf{A}) = 0$, where $\textbf{A}$ is
\begin{equation}
    \begin{split}
    \mathbf{A} =
        \begin{pmatrix}
        -k_z^2+\frac{\omega^2}{c^2}\epsilon_{11} & \frac{\omega^2}{c^2}\epsilon_{12} & k_x k_z +\frac{\omega^2}{c^2}\epsilon_{13}\\
        \frac{\omega^2}{c^2} \epsilon_{21} & -k_x^2-k_z^2+\frac{\omega^2}{c^2}\epsilon_{22} & \frac{\omega^2}{c^2}\epsilon_{23}\\
         k_x k_z +\frac{\omega^2}{c^2}\epsilon_{31} &\frac{\omega^2}{c^2}\epsilon_{32} & -k_x^2+\frac{\omega^2}{c^2}\epsilon_{33}
        \end{pmatrix},
    \end{split}
    \label{eq:aematrix}
\end{equation}
and $n_{e,i}=\sqrt{\epsilon_{ii}}$ are
\begin{equation}
\begin{aligned}
    n_{e,1}(\Omega t, \gamma)&\simeq n_e+\delta n \cos^2(\gamma)\left[\sin^2(\Omega t)+\tan^2(\gamma)\right],
    \\n_{e,2}(\Omega t, \gamma)&\simeq n_o-\delta n \cos^2(\gamma)\sin^2(\Omega t),
    \\n_{e,3}(\Omega t, \gamma)&\simeq n_o-\delta n\sin^2(\gamma).
    \label{eq:nii}
\end{aligned}
\end{equation}

\subsubsection{Deriving nonlinear Schrödinger equations}
We model light propagation starting from the vector wave equation to derive a nonlinear Schrödinger equation (NLSE) for the o- and e-beam by substituting the electric field from Eq.~\eqref{eq:efield} and eigenvalue solutions for the wave vectors in Eq.~\eqref{eq:nii}. We assume $\nabla\cdot\mathbf{E}\simeq0$, $v\ll c$ and weak birefringence to obtain two NLSEs for the o- and e-beams

and applying the slowly varying envelope approximation, we obtain
\begin{equation}
\begin{aligned}
2\imath\mathbf{k}_o\cdot\nabla A+\nabla_{\perp}^2A+\left(\frac{\omega^2}{\epsilon_0c^2}\hat{\mathbf{E}}_o\cdot\mathbf{\epsilon}\hat{\mathbf{E}}_o-\mathbf{k}_o^2\right)A=0,
\\2\imath\mathbf{k}_e\cdot\nabla B+\nabla_{\perp}^2B+\left(\frac{\omega^2}{\epsilon_0c^2}\hat{\mathbf{E}}_e\cdot\mathbf{\epsilon}\hat{\mathbf{E}}_e-\mathbf{k}_e^2\right)B=0,
\label{eq:WE7}
\end{aligned}
\end{equation}
where $\nabla_{\perp}^2=\frac{\partial^2}{\partial x^2}+\frac{\partial^2}{\partial y^2}$ is the transverse Laplacian, and $k_{o,e}^2(\Omega t,\theta)=\frac{n_{o,e}^2(\Omega t,\theta)\omega^2}{c^2}$ are the wave vectors with refractive indices defined as in Eqs.~\eqref{eq:new_ord}, and ~\eqref{eq:new_ext}.

Let us suppose that dispersion is large such that the phase and group indices are significantly different from one another. As a result, the group indices can be written as
\begin{equation}
\begin{aligned}
n_o\rightarrow n_o^{g}&=n_o+\omega_0\left(\frac{\partial n_o}{\partial \omega}\right)_{\omega_0},\\
n_e\rightarrow n_e^{g}&=n_e+\omega_0\left(\frac{\partial n_e}{\partial \omega}\right)_{\omega_0},
\end{aligned}
\label{eq:n_oe_eff}
\end{equation}
where $n_{o,e}^g$ are the group refractive indices for the o- and e-beams, respectively. Nonlinear contributions contribute to large dispersion, with an intensity dependence written as
\begin{equation}
\begin{aligned}
n_{g_{o,e}}\rightarrow n_{g_{o,e}}^{\mathrm{eff}} \approx n_{o,e}^g+\frac{1}{2}n_2^g I_{o,e},
\end{aligned}
\label{eq:n_oe_eff_NL}
\end{equation}
where $n_2^g$ is the nonlinear group index, and $I_{o,e}$ are the intensities for both the o- and e-beams, respectively. It is important to note that we assume that both beams affect the other equally as the self-action. We can define an effective refractive index for the o- and e-beams

\begin{equation}
\begin{aligned}
&n_o^{\mathrm{eff}^2}=n_o+\omega_0 \left(\frac{\D n_o}{\D \omega}\right)_{\omega_0}+n_2^g|A|^2/2,\\&
n_e^{\mathrm{eff}^2}=n_e+\omega_0 \left(\frac{\D n_e}{\D \omega}\right)_{\omega_0}+n_2^g|B|^2/2.
\end{aligned}
\end{equation}

We must also apply a Lorentz transformation, $\Delta k$, to be in a moving reference frame, 
\begin{equation}
  \Delta k =-\frac{\omega_0\Omega r}{c^2}.
  \label{lorentztrans}
\end{equation}
Applying the transform and substitutions for the effective refractive indices, the transverse beam profile becomes 
\begin{equation} 
\begin{aligned}
 &A = a(x,y,z) \exp\Bigg({\imath \frac{\left(n_o^{\mathrm{eff}^2}-n_o^2\right)}{2k_o^2c^2}\omega_0^2\mathbf{\Delta k_o'}\cdot\hat{r}}\Bigg),\\& B = b(x,y,z) \exp\left({\imath\frac{\left(n_e^{\mathrm{eff}^2}-n_e^2\cos^2\left(\gamma\right)\right)}{2k_e^2\cos^2\left(\gamma\right)c^2}\omega_0^2\mathbf{\Delta k_e'}\cdot\hat{r}}\right).
\end{aligned}
\end{equation}
Substituting the fields into the two generalized coupled NLSEs, we arrive at
\begin{widetext}
\begin{equation}
\begin{aligned}
&\partial_z a=\frac{\imath}{2k_o}\nabla_{\perp}^2a+\frac{\imath k_o}{n_o}\Delta n_{NL}a-\frac{\omega_0^2 \left(n_o^{\mathrm{eff}^2}-n_o^2\right)\Delta k}{2k_o^3c^2}\partial_y a,
\\ \partial_z b=\frac{\imath}{2k_e\cos^2(\gamma)}\nabla_{\perp}^2b+&\frac{\imath k_e}{n_e\cos^2(\gamma)}\Delta n_{NL}b+2\tan(\gamma)\left(\cos(\Omega t)\partial_x b +\sin(\Omega t)\partial_y b\right)-\frac{\omega_0^2 \left(n_e^{\mathrm{eff}^2}-n_e^2\cos^2(\gamma)\right)\Delta k}{2k_e^3\cos^4(\gamma)c^2}\partial_y b.    
\end{aligned}
\label{eq:WE_a}
\end{equation}
where $\Delta n_{NL}(I)$ is a nonlinear correction term, and $I$ is the total beam intensity. This contribution becomes relevant with intense illumination, further discussed in Sec.~\ref{sec:3}. Assuming monochromatic light, weak birefringence ($\delta n\ll1$) and $v\ll c$, the coupled nonlinear Schrödinger equations become
\begin{equation}
\begin{aligned}
 &\partial_z a=\frac{\imath}{2k_o}\nabla_{\perp}^2a-\frac{\imath k_o}{n_o}\Delta n_{NL}a- \frac{n_{g,o}^{NL}}{c}\partial_y a,
\\\partial_z b=\frac{\imath}{2k_e\cos^2(\gamma)}\nabla_{\perp}^2b&+\frac{\imath k_e}{n_e\cos^2(\gamma)}\Delta n_{NL}b+2 \tan(\gamma)\left(\cos(\Omega t)\partial_x b +\sin(\Omega t)\partial_y b\right)-\frac{n_{g,e}^{NL}}{c}\partial_y b.  
\end{aligned}
\label{eq:WE_nga}
\end{equation}
\end{widetext}
where 
\begin{equation}
    n_{g,o}^{NL}=\frac{\omega_0^2 \left(n_o^{\mathrm{eff}^2}-n_o^2\right)\Delta k}{2k_o^3c^2},
\end{equation}
and
\begin{equation}
    n_{g,e}^{NL}=\frac{\omega_0^2 \left(n_e^{\mathrm{eff}^2}-n_e^2\cos^2(\gamma)\right)\Delta k}{2k_e^3\cos^4(\gamma)c^2}.
\end{equation}


\subsection*{Derivation of transverse shift due to nonlinear photon drag}

Due to the motion of the medium, we suppose that a small deviation impinges on the beam in the transverse plane at an angle $\theta$. We find the angle $\theta$ as a function of phase index $n_0$, and the group index $n_g$
\begin{equation}
\tan\theta=\frac{v}{c}\left(n_g-\frac{1}{n_0}\right),
\label{eq:dragangle}
\end{equation}
where $v$ is the medium speed, and $c$ is the speed of light. We can find the amount of transverse shift $\Delta y$ in terms of the medium length L, replacing $\tan(\theta)=y/L$ to find 
\begin{equation}
\Delta y=\frac{vL}{c}\left(n_g-\frac{1}{n_0}\right).
\label{eq:lindrag}
\end{equation}

We write the phase index in terms of the complex dielectric permittivity $\epsilon$, such that $n_0=\sqrt{\epsilon}$. The permittivity is written as
\begin{equation}
\epsilon(\omega)=\epsilon_r(\omega)+\imath\epsilon_i(\omega),
\label{eq:eps1}
\end{equation}
where $\epsilon_r(\omega)$ and $\epsilon_{i}(\omega)$ are the real and imaginary parts, respectively. However, we have assumed that the medium is lossless. Therefore, we can neglect the contribution of the imaginary part of the permittivity such that
\begin{equation}
n_0=\sqrt{\epsilon_r}.
\label{eq:phaseindex}
\end{equation}
Since it is the difference in phase and group indices that shows the importance of the drag effect, the imaginary part plays a negligible role in the drag and, therefore, can be omitted.

Knowing that $n_g=n_0+\omega \frac{\D n_0}{\D \omega}$, we can write the group index in terms of the permittivity as  
\begin{equation}
n_g=\sqrt{\epsilon_r(\omega_0)}+\frac{\omega_0}{2\sqrt{\epsilon_r(\omega_0)}}\left(\frac{\D\epsilon_r}{\D\omega}\right)_{\omega_0},
\label{eq:groupindex}
\end{equation} 
and substitute Eqs.~\eqref{eq:phaseindex} and ~\eqref{eq:groupindex} into Eq.~\eqref{eq:lindrag}, and find the linear photon drag effect
\begin{equation}
\begin{aligned}
    \Delta y=\frac{v L}{c}\left[\epsilon_r(\omega_0)+\frac{\omega_0}{2 \sqrt{\epsilon_r(\omega_0)}}\frac{d\epsilon_r}{d\omega}-\frac{1}{\sqrt{\epsilon_r(\omega_0)}}\right].
\end{aligned}
\label{eq:lindrageps}
\end{equation}

In the presence of an intense laser beam, certain media can exhibit large group indices ($n_g\approx10^6$~\cite{piredda2007slow}). Using these media, the linear photon drag effect can be extended to be nonlinear. 
Firstly, we consider the lowest order correction to the permittivity in the form of $\Delta n$, where the correction is an instantaneous Kerr-type nonlinearity of the form, $\Delta n(\omega) = n_2(\omega)I$. Here, $n_2$ is the nonlinear refractive index coefficient, and $I$ is the intensity of the input beam. Assuming $\Delta n\leq n_0$, we find
\begin{equation}
  \epsilon_r(\omega_0) = \left( n(\omega) +\Delta n(\omega_0)\right)^2 \simeq n(\omega_0)^2+2 n(\omega_0) \Delta n(\omega_0).
  \label{eq:epsperturb}
\end{equation}
From the group index contribution in Eq.~\eqref{eq:lindrageps}, the derivative of the permittivity is also needed. Therefore, we derive the permittivity with respect to the frequency and evaluate at $\omega=\omega_0$ to find
\begin{equation}
\left(\frac{\D\epsilon_r}{\D\omega}\right)_{\omega_0}\simeq2n_0\left[\left(\frac{\D n}{\D\omega}\right)_{\omega_0}
+I\left(\frac{\D \Delta n}{\D\omega}\right)_{\omega_0}\right].
\label{eq:n2omega}
\end{equation}

Subbing Eqs.~\eqref{eq:epsperturb} and ~\eqref{eq:n2omega} into Eq.~\eqref{eq:groupindex}, we can find a nonlinear group index $n_g^{NL}$
\begin{equation}
n_g^{NL}=n_g^0+n_2^g I,
\label{eq:groupindexfinal}
\end{equation}
where $n_g^0$ is the linear group index, $n_2^g$ is the nonlinear group index written as
\begin{equation}
n_0^g=n_0 +\omega \left(\frac{\D n_0}{\D\omega}\right)_{\omega_0},
\end{equation}
and 
\begin{equation}
n_2^g=I^{-1}\left(\Delta n +\omega \left(\frac{\D \Delta n}{\D\omega}\right)_{\omega_0}\right).
\end{equation}
Using the $n_g^{NL}$, we can define a nonlinear angle
\begin{equation}
\tan \theta_{NL}=\frac{v}{c}\left(n_g^{NL}-\frac 1{n_0}\right),
\end{equation}
which can be described by $\theta_{NL}=\theta\pm\Delta\theta$ depending on the nonlinear response of the medium. With this nonlinear angle, we exploit again the geometry of the system to find a transverse shift due to a nonlinear equivalent of the photon drag effect 
\begin{equation}
\Delta{y_{NL}} = L \tan(\theta_{NL}) = \frac{L v}{c}\left(n_{g}^{NL}+\frac{1}{n_0}\right).
\label{eq:dragnonlin}
\end{equation}

Under certain conditions, systems have a large nonlinear group index, and therefore we can suppose $|n_g^{NL}|\gg n_0$. Depending on the dispersion, the system could exhibit normal $n_g^{NL}>0$ or anomalous $n_g^{NL}<0$ dispersion and the drag would be negative and therefore considered slow or fast light, respectively. 
As a result, the nonlinear transverse shift in position scales linearly with the $n_g^{NL}$
\begin{equation}
\Delta{y_{NL}}\approx \frac{Lv}{c}n_{g}^{NL}.
\label{eq:dragnonlinapprox}
\end{equation}
\subsection*{Rotation Speed and Intensity Dependence of Nonlinear Group Index}
Our model has considered an instantaneous nonlinearity, which is a simplification of the system in question. The rotation speed makes the nonlinearity act as a non-instantaneous response, particularly dictating the rates, and, therefore, a temporal response. Fast timescales can act and locally affect the beam when considering slow speeds, while fast speeds are associated with long-lived effects on integer multiple full crystal rotations. In the following two sections, we will describe how the thermal and optical nonlinear response contribute to the index gradient that controls the magnitude of the $n_g^{NL}$ and, ultimately, the amount of transverse shift experienced by the optical beam upon propagation.
\subsubsection{Thermal Contribution to $n_g^{NL}$}

\begin{figure}[!t]

\includegraphics[width=0.85\columnwidth]{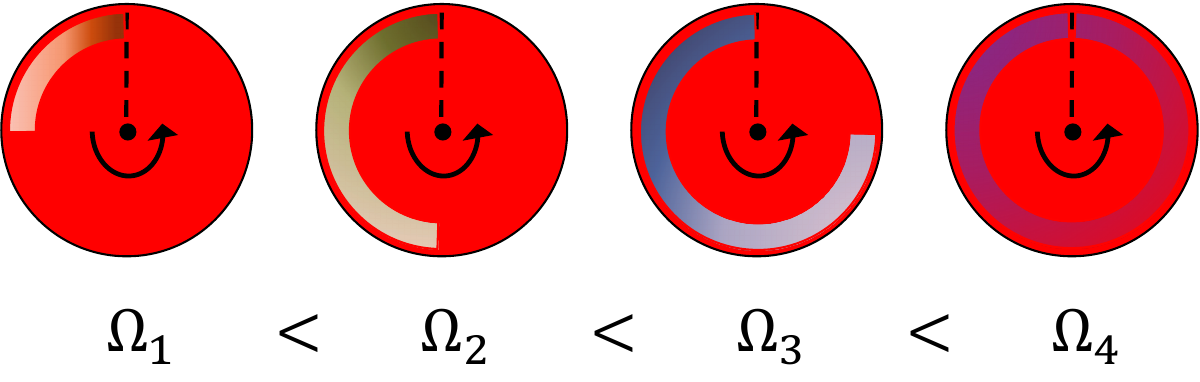}
\caption{
{\bf Schematic optical beam distributing heat over a rotating ruby rod crystal face.} The heat distribution circles about the crystal cause an index gradient with varying magnitude depending on speed. The heat reaches different distances depending on the rotation speed until it reaches fully around the crystal, producing a smaller magnitude of index gradients at higher rotation speeds.}
\label{fig:10}
\end{figure}

\begin{figure*}
\centering
\includegraphics[width=0.99\textwidth]{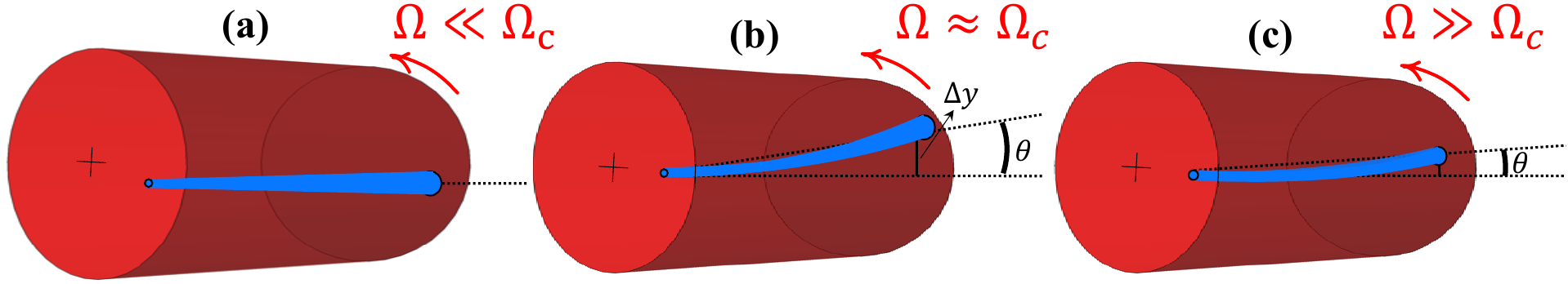}
\caption{
{\bf Schematic of curved trajectory induced by a moving index gradient created by the nonlinear refraction.} Part \textbf{(a)} shows the trajectory at very slow speeds, \textbf{(b)} shows the largest transverse shift around $\Omega=\Omega_c$, and \textbf{(c)} shows a smaller transverse shift at high speeds.}
\label{fig:11}
\end{figure*}

When considering slower rotation speeds, the thermal nonlinear response contributes most to $n_g^{NL}$ acting typically on the order of several hundred microseconds~\cite{doiron2004time}. A depiction of the heat deposition and distribution due to the thermal nonlinear response is shown in Fig. \ref{fig:10}. Sufficiently slow speeds generate a relatively constant index gradient in magnitude. The index gradient steadily increases the transverse shift with an increase in rotation speed up to characteristic speed, $\Omega_c$, and decreases thereafter due to less index gradient. Thus, a diminishing contribution from the thermal nonlinear response with increasing rotation speed is observed up to a steady-state response like in the case of $\Omega_4$ in Fig. \ref{fig:10}. The characteristic speed is related to the timescale of the interaction as $\Omega_c= (2\pi\tau_c)^{-1}$; $\tau_c$ is a characteristic interaction time. The timescale for the thermal nonlinear response is associated with an exponential decay that reaches a maximum index gradient as 
\begin{equation}
\Delta n_{\mathrm{max}}^g(t)=n_2^g I_0 e^{-(t-t_0)/\tau_c},
\label{eq:ngmax}
\end{equation}
where $I_0$ is the input intensity, and $t_0$ is an offset time. We rewrite Eq.~\eqref{eq:ngmax} as a function of the rotation speed, $(t-t_0)/\tau_c=(\Omega-\Omega_0)/\Omega_c$, where $\Omega_c$ is a characteristic rotation speed, and $\Omega_0$ is an offset rotation speed, such that 
\begin{equation}
n_g^{NL}= n_g^0 + n_2^g I = n_g^0+n_2^g I_0 e^{-(\Omega-\Omega_0)/\Omega_c}.
\label{equation44}
\end{equation}
The $n_g^{NL}$ in the form of Eq.~\eqref{equation44} shows the general case when considering the system dynamics. For $\Omega<\Omega_c$, the nonlinear response is predominantly thermal, denoted $\Omega_f=(2\pi\tau_f)^{-1}$, where $\tau_f\approx200$ $\mu$s~\cite{doiron2004time}. The thermal contribution to the $n_g^{NL}$ is therefore
\begin{equation}
n_g^{NL}= n_g^0+n_{2,\mathrm{therm}}^g I_0,
\label{ngtherm}
\end{equation}
where $n_{2,\mathrm{therm}}^g$ is the thermal nonlinear group index,
\begin{equation}
n_{2,\mathrm{therm}}=-\alpha_f n_2^g e^{-(\Omega-\Omega_0)/\Omega_f},
\end{equation}
and $\alpha_f$ is a phenomenological scaling factor.

For $\Omega>\Omega_c$, the optical nonlinear response take over, denoted $\Omega_s=(2\pi\tau_s)^{-1}$. Some optical processes depend on the lifetime of the excited state of an atom, which can be on the order of $\tau_s=3\sim5$ ms~\cite{piredda2007slow} that fit into this region. If the effect was only a thermal nonlinear process, the shift would progressively approach zero with higher rotation speeds and so we must consider both optical and thermal responses.

\subsubsection{Optical Contribution to $n_g^{NL}$}
In order to facilitate the explanation of the optical contributions, we'll begin using experimental parameters here. We input a Gaussian beam profile in the $xy$ plane with a beam waist of 10 $\mu$m at a position $x_0=0.8R$, where $R$ is the radius ($R = 0.35 $ cm) and $L$ is the length ($L=2$ cm) of the crystal, respectively. In the regime of $\Omega>\Omega_c$, $n_g^{NL}$ is the optical nonlinear response of CPO, giving rise to large group indices~\cite{piredda2007slow}. Similarly, we write the optical response of $n_g^{NL}$
\begin{equation}
n_g^{NL}= n_g^0+n_{2,\mathrm{opt}}^g I_0,
\label{ngopt}
\end{equation}
where the optical nonlinear group index takes the form
\begin{equation}
n_{2,\mathrm{opt}}=\alpha_s n_2^g e^{-(\Omega-\Omega_0)/\Omega_s},
\end{equation}
where $\alpha_s$ is a scaling constant. We note the sign of the nonlinear contribution is now positive. Certain systems exhibit a large negative group index $n_g\approx-10^6$~\cite{https://doi.org/10.48550/arxiv.2301.13300}, and the value was set to be $n_2^g I_0=-1.1\times10^6$. The coexistence of the two nonlinear processes results in a purely positive transverse shift at the output of the crystal. $n_2^g I_0$, however, can be set to any value depending on the system at hand.

Since the system impinges an index gradient that moves with the medium, the trajectory can be curved, contributing to the overall transverse shift. An example of a curved trajectory is shown in Fig. \ref{fig:11}, which conveys the idea of a non-zero output angle at the end of the medium. It is crucial to measure this output angle to distinguish if nonlinear deflection has also contributed to the transverse shift. The photon drag effect generally exits the crystal parallel to the input beam. However, we stress that although the beam exits at an angle that is not parallel, it propagates in a straight line from that output angle, as expected in free space.

\subsection*{The effect of thermal nonlinear response on the dielectric tensor \label{sec:3}}

Although the index gradient, $\Delta n_{NL}'$, impinged by a thermal nonlinearity, does not contribute strongly to the amount of transverse shift, analyzing the effect is worth noting. Let's consider intense illumination impinged on media with nonlinear refraction, creating an index gradient modifying the dielectric tensor. Therefore, for first-order correction in the crystal frame, the dielectric permittivity is 
\begin{equation}
\mathbf{\epsilon}'\longrightarrow\mathbf{\epsilon}'+\Delta\mathbf{\epsilon}_{NL}',
\end{equation}
where $\Delta \mathbf{\epsilon}_{NL}'$ is written as
\begin{equation}
\Delta\mathbf{\epsilon}_{NL}'=2\epsilon_0\Delta n_{NL}'\left(\begin{array}{ccc}
n_e & 0 & 0\\
0 & n_o & 0\\
0 & 0 & n_o
\end{array}\right).
\label{eq:crystal_epsNL}
\end{equation}
The index gradient created by nonlinear refraction is also affected by the rotation of the medium and tilt angle $\gamma$, where 
\begin{equation}
\Delta n_{NL}'\rightarrow \Delta n_{NL}'(\Omega t,\gamma).
\end{equation}

In thermal media~\cite{marcucci2019optical}, the index gradient due to thermal nonlinear response is 
\begin{equation}
\Delta n_{NL}' =\left(\frac{\partial n}{\partial T}\right)_0 \Delta T(\mathbf{r}'),
\label{eq:n_T}
\end{equation}
where $\left(\frac{\partial n}{\partial T}\right)_0$ is the medium's thermo-optic coefficient at thermal equilibrium (steady-state response) and $\Delta T(\mathbf{r}')$ is the temperature variation about the point $\mathbf{r}'=(x',y',z')$. $\Delta T(\mathbf{r}'$) for a stationary medium is governed by the $3$D heat equation
\begin{equation}
\left(\partial^2_{x'}+\partial^2_{y'}+\partial^2_{z'}\right)\Delta T(\mathbf{r}')=-\gamma|\mathbf{E}'(\mathbf{r}')|^2,
\label{eq:temperature0}
\end{equation}
with $\gamma_l=(L_{loss}\rho_0c_P D_T)^{-1}$, where $L_{loss}$ is the loss characteristic length, $\rho_0$ the material density, $c_P$ the specific heat at constant pressure, and $D_T$ is the thermal diffusivity. The solution can be written in terms of a Green function $G(\mathbf{r}')$, which depends only on the sample geometry and the boundary conditions and expresses the nonlocality of the nonlinear effect
\begin{equation}
\Delta T(\mathbf{r}') =\iiint \mathrm{d}\tilde{\mathbf{r}}'G'(\mathbf{r}'-\tilde{\mathbf{r}}')|\mathbf{E}'(\tilde{\mathbf{r}}')|^2.
\label{eq:temperature}
\end{equation}

Around the neighborhood of the medium's midpoint, i.e. $z_0=L/2$, in the longitudinal parabolic approximation of characteristic width $L_{nloc}=\sqrt{\frac{|n_2|}{\gamma_l\left|\frac{\partial n}{\partial T}\right|_0}}\propto\sqrt{L_{loss}}$~\cite{marcucci2019optical}, $n_2$ the nonlinear refractive index, $L_{loss}$ is the characteristic loss length, Eq.~\eqref{eq:temperature} reads
\begin{equation}
\Delta T_{\perp}(\mathbf{r_{\perp}}')=\iint \mathrm{d}\tilde{\mathbf{r}}_{\perp}'G_{\perp}'(\mathbf{r_{\perp}}'-\tilde{\mathbf{r}}_{\perp}')I_{\perp}'(\tilde{\mathbf{r}}_{\perp}'),
\label{eq:temperature2}
\end{equation}
\begin{widetext}
\noindent with $I_{\perp}'(\mathbf{r_{\perp}}')=\frac1L\int \mathrm{d}z'I'(\mathbf{r_{\perp}}',z')$, $I'(\mathbf{r}')=|\mathbf{E}'(\mathbf{r}')|^2=|\mathbf{E}(\mathbf{r})|^2=I(\mathbf{r})$, and $\mathbf{r_{\perp}}'=(x',y')$.
Assuming absorption is low ($L\ll L_{loss}$), we find $\Delta T(\mathbf{r}') \sim \Delta T_{\perp}(\mathbf{r_{\perp}}')$ and $\partial_{z'} I'(\mathbf{r}')\sim 0$. As a result, the index gradient impinged on the crystal by the thermal nonlinear response is
\begin{equation}
\begin{aligned}
\Delta n_{NL}'(\mathbf{r_{\perp}}')=n_2 \iint \mathrm{d}\tilde{\mathbf{r}}_{\perp}'K'(\mathbf{r_{\perp}}'-\tilde{\mathbf{r}}_{\perp}')I'(\tilde{\mathbf{r}}_{\perp}')- n_{o,e},
\label{eq:nonlinearity}
\end{aligned}
\end{equation}
with $n_2K'(\mathbf{r_{\perp}})=\Bigl(\frac{\partial n}{\partial T}\Bigr)_0G_{\perp}'(\mathbf{r}_{\perp}')$. The index gradient is now written as

\begin{equation}
\Delta n_{NL}(x,y,\Omega t,\gamma)= n_2 \iint \mathrm{d}\widetilde{x} \mathrm{d}\widetilde{y} K_\gamma \left(\Delta x, \Delta y, \Omega t \right) I(\widetilde{x},\widetilde{y}) - n_{o,e}, 
\label{eq:nnl}
\end{equation}
where $K_\gamma$ is the nonlinear nonlocal kernel function affected by the weak birefringence, written as
\begin{equation}
K_\gamma \left(x, y, \Omega t, \gamma\right)= K' \left[\cos(\gamma)\left(\cos(\Omega t) x+\sin(\Omega t) y\right), -\sin(\Omega t) x+ \cos(\Omega t) y\right].
\end{equation}

Understanding the importance of the kernel function further requires a definition in Fourier space, as the NLSE adds the nonlinear response as a phase term. Therefore, the kernel function in Fourier space is 
\begin{equation}
K_\gamma\left(k_x, k_y, \Omega t,\gamma\right)=\frac{1}{2\pi\left[(k_x')^2+(k_y')^2\right] \left[1+L_{\mathrm{nloc}}^2\left(1-\exp\left(-\frac{t}{\tau_N}\right)\right)\right]},
\end{equation}
where $L_{\mathrm{nloc}}$ is the nonlocal length, $\tau_N$ is the non-instantaneous timescale, and $k_x'$ and $k_y'$ are the $x$ and $y$ wave vectors in the crystal reference frame. The nonlocality does not play a huge role but scales the non-instantaneous response. The non-instantaneous part only plays a role when the timescales of the interactions are long-lived, on the order of seconds or more. A summary of the variables and their functionality can be found in Table \ref{table3}. With all relevant variables, we can define how the dielectric permittivity is affected by an index gradient. Therefore in the lab frame, we write the dielectric tensor with all perturbative terms under the assumption of weak birefringence
\begin{equation}
\mathbf{\epsilon}=\mathbf{\epsilon}'+\mathbf{\epsilon}''(\gamma,\Omega t)+\Delta\mathbf{\epsilon}_{NL}'(x,y,\Omega t,\gamma).
\end{equation}

\begin{table}[t!]
\centering
\begin{tabular}{|p{2cm}|p{5cm}|}
\hline
Variable & Functionality\\
\hline\hline
$K_\gamma$&Thermal Kernel Function\\
\hline
$\left(\frac{\partial n}{\partial T}\right)_0$ & Thermo-optic coefficient\\
\hline
$\Delta T(\mathbf{r}')$& Temperature Variation\\
\hline
$L_{loss}$& Characteristic Loss Length\\
\hline
$\rho_0$ & Material Density\\
\hline
$c_P$&Specific Heat at Constant Pressure\\
\hline
$D_T$&Thermal Diffusivity\\
\hline
$G(\textbf{r'})$& Green's Function\\
\hline
$L_{nloc}$ & Nonlocal Length\\
\hline
$t_N$ & Non-instantaneous timescale\\
\hline
\end{tabular}
\caption{\textbf{Summary of the relevant variables used to calculate the index gradient $\Delta n_{NL}$ due to a thermal nonlinearity.} The definition of the variables is used to clarify the function of each variable within the derivative.}
\label{table3}
\end{table}

\subsection*{Simulations and Split-Step Fourier Method \label{sec:sims}}
We apply the Split-Step Fourier Method to simulate nonlinear propagation through the rotating ruby rod of length $L$. We input a Gaussian beam profile in the $xy$ plane focused to a position far from the centre of rotation. The NLSEs in Eq.\eqref{eq:WE_nga} are propagated with the usual formalism of a linear propagator $\hat{D}$ and a nonlinear propagator $\hat{N}$. We can represent the NLSEs in the SSFM formalism as 
\begin{equation}
\begin{aligned}
    &\partial_z a = (\hat{D_o}+\hat{N_o})a,
    \\&\partial_z b = (\hat{D_e}+\hat{N_e})a,
\end{aligned}
\end{equation}
where the linear propagators are defined as
\begin{equation}
\begin{aligned}
    &\hat{D_o}=\frac{\imath}{2k_o}\nabla_{\perp}^2- \frac{n_{g}^{NL}}{c}\partial_y,
    \\&\hat{D_e}=\frac{\imath}{2k_e\cos^2(\gamma)}\nabla_{\perp}^2+\frac{n_{g}^{NL}}{c}\partial_y+2 \tan(\gamma)\left(\cos(\Omega t)\partial_x  +\sin(\Omega t)\partial_y \right),
\end{aligned}
\end{equation}

and the nonlinear propagators are defined as
\begin{equation}
\begin{aligned}
    &\hat{N_o}=\frac{\imath k_o}{n_o}\Delta n_{NL},\\&\hat{N_e}=\frac{\imath k_e}{n_e\cos^2(\gamma)}\Delta n_{NL}.
\end{aligned}
\end{equation}

The fields $a$ and $b$ represent the o- and e- fields, respectively. We can interchange between real space ($a$ and $b$) and Fourier space ($\hat{a}$ and $\hat{b}$) using a Fourier Transform (FT), or vice versa with the inverse FT, that is

\begin{equation}
\begin{aligned}
    &a(x,y,z)=\frac{1}{2\pi}\iint_{\mathbb{R}^2}dk_xdk_y \hat{a}\left(k_x,k_y,z\right)e^{-\imath\left(k_x x+k_y y\right)},\\&    \hat{a}(k_x,k_y,z)=\frac{1}{2\pi}\iint_{\mathbb{R}^2}dxdy a\left(x,y,z\right)e^{\imath\left(k_x x+k_y y\right)}.
\end{aligned}
\end{equation}
where $b$ and $\hat{b}$ are written in a similar fashion.

We apply the linear propagators in Fourier space between the points $z$ and $z+h$, that is 
\begin{equation}
    \begin{aligned}
        &\exp\left(\frac{ h}{2}\hat{D_o}\right)=\exp\left(\frac{-\imath h}{2}\left[\frac{-1}{2k_o}(k_x^2+k_y^2)-\imath \frac{n_{g}^{NL}}{c}k_y\right]\right),\\&\exp\left(\frac{h}{2}\hat{D_e}\right)=\exp\Bigg(\frac{-\imath h}{2}\Bigg[\frac{-1}{2k_e\cos^2(\gamma)}(k_x^2+k_y^2)-\imath \frac{n_{g}^{NL}}{c}k_y+2 \tan(\gamma)\left(\cos(\Omega t)k_x +\sin(\Omega t)k_y \right)\Bigg]\Bigg).
    \end{aligned}
\end{equation}
\end{widetext}
The linear step is then applied by taking the inverse FT of the product of the linear propagator in Fourier space and the FT of the field
\begin{equation}
    \begin{aligned}
        &K_{1n}=h f(x_n,y_n),    \\&K_{2n}=h f(x_n+\frac{1}{2}h, y_n+\frac{1}{2}K_{1n}),\\&y_{n+1}=y_n + K_{2n}+O(h^3).
    \end{aligned}
\end{equation}

We then will apply the nonlinear propagators assuming the boundary conditions $y'=f(x,y)$ and $y(x_o)=y_o$. At the n$^{\mathrm{th}}$ step, we have 
\begin{equation}
    \begin{aligned}
        &a=\mathrm{IFT}\left[\exp\left(\frac{ h}{2}\hat{D_o}\right)\mathrm{FT}\left[a\right]\right],\\&b=\mathrm{IFT}\left[\exp\left(\frac{ h}{2}\hat{D_e}\right)\mathrm{FT}\left[b\right]\right].
    \end{aligned}
\end{equation}

Recall that $\Delta n_{NL}$ is dependent on both fields, that is $\Delta n_{NL}:=\Delta n_{NL}\left(|a|^2+|b|^2\right)$. Therefore, the nonlinear propagators are functions of both $a$ and $b$. We apply the propagators in the following manner
\begin{equation}
    \begin{aligned}
        &A_{1n}=h \hat{N_o}\left(a_n,b_n\right)a_n,\\&B_{1n}=h \hat{N_e}\left(a_n,b_n\right)b_n,\\&A_{2n}=h \hat{N_o}\left(a_n+\frac{1}{2}A_{1n},b_n+\frac{1}{2}B_{1n}\right)\left(a_n+\frac{1}{2}A_{1n}\right),
        \\&B_{2n}=h \hat{N_e}\left(a_n+\frac{1}{2}A_{1n},b_n+\frac{1}{2}B_{1n}\right)\left(b_n+\frac{1}{2}B_{1n}\right),\\&a_{n+1}=a_n+A_{2n},\\&b_{n+1}=b_n+B_{2n}.
    \end{aligned}
\end{equation}

Once we have applied these propagators over the entire length of the crystal, all transverse movement at the output crystal facet can be tracked. Furthermore, we can take the average positions of the transverse trajectories, weighted by the intensities of these trajectories, to extract the transverse shift, which is controlled by the rotation speed input intensity.


\end{document}